\documentclass[twocolumn]{aastex631}
\usepackage{textgreek}
\usepackage{needspace}
\received{May 13, 2022}
\revised{June 28, 2022}
\accepted{June 29, 2022}

\submitjournal{ApJ}

\shorttitle{The spindown of fully-convective M dwarfs}
\shortauthors{Pass et al.}
\graphicspath{{./}{figures/}}

\begin{document}

\title{Constraints on the Spindown of Fully-Convective M Dwarfs Using Wide Field Binaries}

\widowpenalty=10

\author[0000-0002-1533-9029]{Emily K. Pass}
\affiliation{Center for Astrophysics $\vert$ Harvard \& Smithsonian, 60 Garden Street, Cambridge, MA 02138, USA}

\author[0000-0002-9003-484X]{David Charbonneau}
\affiliation{Center for Astrophysics $\vert$ Harvard \& Smithsonian, 60 Garden Street, Cambridge, MA 02138, USA}

\author{Jonathan M. Irwin}
\affiliation{Center for Astrophysics $\vert$ Harvard \& Smithsonian, 60 Garden Street, Cambridge, MA 02138, USA}
\affiliation{Institute of Astronomy, University of Cambridge, Madingley Road, Cambridge CB3 0HA, UK}

\author[0000-0001-6031-9513]{Jennifer G. Winters}
\affiliation{Center for Astrophysics $\vert$ Harvard \& Smithsonian, 60 Garden Street, Cambridge, MA 02138, USA}



\begin{abstract}

M dwarfs remain active over longer timescales than their Sunlike counterparts, with potentially devastating implications for the atmospheres of their planets. However, the age at which fully-convective M dwarfs transition from active and rapidly rotating to quiescent and slowly rotating is poorly understood, as these stars remain rapidly rotating in the oldest clusters that are near enough for a large sample of low-mass M dwarfs to be studied. To constrain the spindown of these low-mass stars, we measure photometric rotation periods for field M dwarfs in wide binary systems, primarily using TESS and MEarth. Our analysis includes M-M pairs, which are coeval but of unknown age, as well as M dwarfs with white dwarf or Sunlike primaries, for which we can estimate ages using techniques like white dwarf cooling curves, gyrochronology, and lithium abundance. We find that the epoch of spindown is strongly dependent on mass. Fully-convective M dwarfs initially spin down slowly, with the population of 0.2--0.3M$_\odot$ rapid rotators evolving from $P_{\rm rot} < 2$ days at 600 Myr to $2 < P_{\rm rot} < 10$ days at 1--3 Gyr before rapidly spinning down to long rotation periods at older ages. However, we also identify some variability in the spindown of fully-convective M dwarfs, with a small number of stars having substantially spun down by 600 Myr. These observations are consistent with models of magnetic morphology-driven spindown, where angular momentum loss is initially inefficient until changes in the magnetic field allow spindown to progress rapidly.
\end{abstract}

\section{Introduction} \label{sec:intro}
Motivated by the discovery of the relatively long-lived ``$C$-sequence" of rapidly-rotating Sunlike stars \citep{Barnes2003}, the Metastable Dynamo Model \citep[MDM;][]{Brown2014}  posits that young stars are born with magnetic dynamos that are weakly-coupled to the stellar wind and experience little spindown when young; stochastically, the dynamo changes to a strongly-coupled mode that rapidly slows the star's rotation, with a characteristic timescale that is mass dependent. As M dwarfs with masses $<$0.35M$_\odot$ are fully convective \citep{Chabrier1997} and therefore lack tachoclines---thought to be significant in generating the magnetic fields of Sunlike stars---one might expect differing behavior for the magnetic dynamos of low-mass M dwarfs; however, the rotation--activity correlation persists in this stellar mass regime \citep[][]{Kiraga2007}. This behavior is observed in a variety of activity metrics, including X-ray emission \citep{Wright2011, Wright2018}, H\textalpha\ luminosity \citep{Newton2017}, UV emission \citep{France2018}, and flares \citep{Medina2020, Medina2022}. The correlation consists of a saturated regime in which activity is uncorrelated with rotation rate and an unsaturated regime where activity correlates with rotation, with the Rossby number of the transition found to be somewhere between 0.1 (from X-rays; \citealt{Wright2018}) and 0.5 (from flares; \citealt{Medina2022}).

While the M dwarf is in the saturated regime, orbiting planets may suffer significant atmospheric loss from these X-rays and flares \citep[e.g.,][]{Lammer2007, Tian2009, Segura2010, Tilley2019, NevesRibeiro2022}. A stochastic component to the epoch of spindown for low-mass stars would therefore have important implications for these planets: some planetary systems would experience prolonged exposure, leaving all their terrestrial worlds denuded. For others, the stellar environment may become quiescent before the atmospheres are lost completely.

Coeval clusters are often used to investigate the time and mass dependencies of stellar spindown. While this technique has been used to study Sunlike stars in clusters as old as 4 Gyr \citep{Barnes2016}, the distance to such clusters is on the order of a kiloparsec and therefore prohibitive for the analysis of faint M dwarfs. A small number of early Ms have been studied in 1--3 Gyr clusters \citep{Agueros2018, Curtis2019, Curtis2020}, but for low-mass M dwarfs, recent studies have been limited to nearby clusters such as Praesepe and the Hyades \citep[][]{Douglas2014, Douglas2016, Douglas2017, Douglas2019} with age estimates in the range 600--800 Myr \citep[e.g.,][]{Choi2016, Cummings2018, gaia2018}. While these works found that most fully-convective M dwarfs are still rapidly rotating at 600 Myr, \citet{Douglas2017} observed a bimodality in the rotation periods of 0.25--0.5M$_\odot$ stars, which they noted may be explained using a modified version of MDM. Notably, the slow rotators in this distribution rotate with intermediate rotation periods (10--30 days). For mid-M dwarfs in the field, \citet{Newton2016} observed slower rotation in the slowly-rotating mode ($P_{\rm{rot}}>70$ days), and noted a dearth of stars with intermediate rotation periods.

As most M dwarfs are still rapidly rotating at 600 Myr, there is a need to probe older populations in order to understand the spindown of these stars. Works such as \citet{Newton2016, Newton2018} studied the rotation rates of an older population of M dwarfs in the field; however, ages for individual field M dwarfs are difficult to determine, as isochrone fitting is uninformative given the long main-sequence lifetimes of these stars. Some works have sidestepped this issue by using galactic kinematics to estimate the ages of populations \citep[e.g.,][]{Newton2016, Lu2021, Medina2022}. We take a different approach by considering a sample of field M dwarfs with widely-separated companions. This method allows us to consider the spindown behavior of individual systems, and therefore, the variability in behavior between systems.

We discuss binary M-M systems in Section~\ref{sec:samp}, where the components share the same unknown age. We then investigate M dwarfs with a hotter wide companion (white dwarfs in Section~\ref{sec:wd}; FGK stars in Section~\ref{sec:fgk-intro}), whose age can be determined by gyrochronology, lithium abundance, or white dwarf cooling rates. We discuss our results in Section~\ref{sec:discussion}, including a comparison with the rotation rates of Praesepe M dwarfs in Section~\ref{sec:pra} and with theoretical spindown models in Section~\ref{sec:garraffo}. We conclude with a summary in Section~\ref{sec:summary}.

\section{M-M Binaries}
\label{sec:samp}

\subsection{Common proper motion search}

We begin our sample selection by conducting a common proper motion (cpm) search in \textit{Gaia} EDR3 \citep{Gaia2016, Gaia2021, Lindegren2021}. We cross-match with the 2MASS catalog \citep{Cutri2003}, discard sources with absolute $K$-band magnitudes outside the range appropriate for M dwarfs ($5 < M_K < 10$ mag), and estimate masses for each remaining source using the \citet{Benedict2016} $K$-band mass-luminosity relation (MLR). As this MLR is defined for masses of 0.08--0.62M$_\odot$, we reject stars outside this range that were not previously removed by our magnitude cut. For computational efficiency, we also restrict our search to sources brighter than $m_R=16$ mag and parallaxes greater than 20 mas (corresponding to distances less than 50pc), with $R$-band magnitudes estimated using the empirical $G-K$ color relation from \citet{Winters2021}.

As we ultimately obtain rotation periods from two different photometric surveys (MEarth/TESS), we create two different proper motion samples based upon the limitations of each program. Given the large pixel scale of TESS (21"), we require that pairs be separated by at least 100"; we use the \texttt{PyAstronomy} routine \texttt{getAngDist} to measure these separations. For our MEarth candidates, we require a separation of only 4". For both samples, we also adopt a maximum separation of 2000" to limit false positives at unphysical separations (although this upper limit can reject some true companions; e.g., \citealt{Mamajek2013}).  We identify cpm pairs using the proper motion ratio and proper motion position angle difference cuts of \citet[][i.e., a threshold of 0.15$^2$ in their Equation 1 and 15\textdegree\ in their Equation 2]{Montes2018}, and also require parallax agreement within 2 mas. This analysis returns 631 M-M pairs with separations 4--2000", of which 129 pairs fall within 100-2000".

\begin{deluxetable*}{llrrrrrr}[t]
\tabletypesize{\footnotesize}
\tablecolumns{7}
\tablewidth{0pt}
 \tablecaption{\textit{Gaia} EDR3 measurements for the common proper motion pairs with TESS rotation periods \label{tab:gaia-tess}}
 \tablehead{
 \colhead{ \vspace{-0.1cm}TIC$_1$} & 
 \colhead{ \vspace{-0.1cm}TIC$_2$} &
 \colhead{$\pi_1$} &
 \colhead{$\pi_2$} &  
 \colhead{$\mu_{\rm RA, 1}$} & 
 \colhead{$\mu_{\rm RA, 2}$} &
 \colhead{$\mu_{\rm DEC, 1}$} &
 \colhead{$\mu_{\rm DEC, 2}$}
 \\
 \colhead{} &
 \colhead{} &
 \colhead{[mas]} &
 \colhead{[mas]} &
 \colhead{[mas/yr]} &
 \colhead{[mas/yr]} &
 \colhead{[mas/yr]} &
 \colhead{[mas/yr]}}
\startdata
22819180 & 22819191 & 42.351 $\pm$ 0.023 & 42.324 $\pm$ 0.036 & 39.221 $\pm$ 0.016 & 37.0793 $\pm$ 0.024 & 26.357 $\pm$ 0.020 & 26.909 $\pm$ 0.030 \\
37664990 & 37664980 & 28.259 $\pm$ 0.025 & 28.307 $\pm$ 0.022 & 250.297 $\pm$ 0.028 & 250.293 $\pm$ 0.026 & -4.559 $\pm$ 0.018 & -3.318 $\pm$ 0.015 \\
43734215 & 43789224 & 28.582 $\pm$ 0.021 & 28.802 $\pm$ 0.025 & -45.377 $\pm$ 0.018 & -45.370 $\pm$ 0.021 & -62.644 $\pm$ 0.019 & -63.478 $\pm$ 0.023 \\
84731806 & 84731362 & 30.266 $\pm$ 0.013 & 30.201 $\pm$ 0.014 & 89.046 $\pm$ 0.012 & 89.946 $\pm$ 0.013 & -2.79 $\pm$ 0.013 & -1.698 $\pm$ 0.014 \\
114953216 & 114985772 & 53.268 $\pm$ 0.033 & 53.354 $\pm$ 0.026 & 358.743 $\pm$ 0.029 & 358.706 $\pm$ 0.026 & -253.35 $\pm$ 0.027 & -254.08 $\pm$ 0.022 \\
197569385 & 197570145 & 23.756 $\pm$ 0.032 & 23.648 $\pm$ 0.042 & 135.137 $\pm$ 0.027 & 135.23 $\pm$ 0.037 & 55.190 $\pm$ 0.024 & 55.853 $\pm$ 0.032 \\
206327797 & 206327810 & 40.077 $\pm$ 0.037 & 40.265 $\pm$ 0.044 & -132.199 $\pm$ 0.019 & -132.002 $\pm$ 0.022 & -23.607 $\pm$ 0.031 & -24.524 $\pm$ 0.038 \\
206617113 & 206617096 & 28.802 $\pm$ 0.016 & 28.769 $\pm$ 0.020 & 1.334 $\pm$ 0.017 & 1.724 $\pm$ 0.019 & 85.134 $\pm$ 0.016 & 84.746 $\pm$ 0.020 \\
256419669 & 52183206 & 101.424 $\pm$ 0.017 & 101.372 $\pm$ 0.048 & 731.088 $\pm$ 0.014 & 730.398 $\pm$ 0.039 & 90.532 $\pm$ 0.017 & 85.967 $\pm$ 0.045 \\
334637014 & 334637029 & 25.691 $\pm$ 0.011 & 25.694 $\pm$ 0.019 & -103.212 $\pm$ 0.013 & -103.625 $\pm$ 0.025 & 131.045 $\pm$ 0.013 & 131.422 $\pm$ 0.023 \\
450297524 & 416857959 & 25.605 $\pm$ 0.018 & 25.627 $\pm$ 0.020 & -36.127 $\pm$ 0.016 & -36.387 $\pm$ 0.018 & 48.756 $\pm$ 0.014 & 47.463 $\pm$ 0.016 
\enddata
\end{deluxetable*}

\vspace{-0.9cm}
\subsection{Refining the TESS sample}
\label{sec:tess}
The Transiting Exoplanet Survey Satellite (TESS) is an ongoing mission to perform space-based photometric monitoring of hundreds of thousands of nearby stars \citep{Ricker2015}. We cross match the \textit{Gaia} EDR3 and DR2 catalogs, allowing us to obtain the TESS Input Catalog (TIC; \citealt{Stassun2019}) identifier for each source by cross matching that catalog with the DR2 identifier. We then use \texttt{Lightkurve} \citep{lightkurve2018} to programmatically identify which sources have short cadence (120s) light curves in the TESS archive. We limit our sample to pairs where both stars have short cadence observations; this criterion removes roughly a third of the preliminary target list.

Next, we visually inspect the light curves and generate Lomb Scargle periodograms for each source, with the goal of identifying which sources exhibit periodic variability. We consider both the Simple Aperture Photometry (SAP) and Pre-search Data Conditioning SAP (PDCSAP) light curves from the MAST archive. While we generally prefer the PDCSAP light curves, as they have had instrumental systematics removed, this process can also remove longer-period rotational modulation. The SAP light curves retain these astrophysical signals, although the instrumental systematics can generate artificial periodicity with a period of half a TESS sector. We restrict our sample to binaries where we observe a periodicity for both components.

We remove a further ten pairs where one or both components have a renormalised unit weight error (RUWE) greater than 2 in \textit{Gaia} EDR3. This value represents the excess noise in the astrometric solution, with an expected value around 1 for a single star. Previous work has found that a large RUWE strongly indicates the presence of an unresolved companion at subarcsecond separation \citep{Vrijmoet2020, Kervella2022}.

We also remove twelve pairs that appear to be single Hyads rather than actual binary systems \citep{Roser2019}; stars in clusters also share similar motions and positions as a result of their formation from the same cloud. Sorting our candidate list by the difference in parallax between components, the rejected Hyads represent twelve out of the thirteen most discrepant sources. Our 2 mas threshold therefore appears to be overly generous given the precision of the \textit{Gaia} parallaxes. The rejected sources also all had large component separations, $\rho$, with the smallest of the group possessing $\rho>800"$. Making these cuts, 14 pairs remain in our TESS sample.

Purity of the sample is key to our analysis; we would rather discard some true binaries than mistakenly include some unassociated systems that could lead us to draw erroneous conclusions about the behavior of coeval pairs. As the \citet{Montes2018} proper motion cuts returned a number of false positives (the Hyads discussed above), we make an additional refinement to our sample by ensuring that the proper motion differences are consistent with a Keplerian orbit using the criterion from \citet[][their Equation 3]{el-Badry2021}, another work studying cpm binaries in \textit{Gaia} EDR3. While their calculation assumes a system mass of 5M$_\odot$, we use the masses estimated from the \citet{Benedict2016} $K$-band relation. All of the Hyads and three of our remaining pairs fail this criterion; notably, the three pairs with the largest angular separation, and hence the pairs that are most likely to be unphysical. These likely unphysical pairs are TIC 106493402/106344480, 273226810/403995704, and 25902832/25837464. We note that TIC 273226810/403995704 was the pair with the large parallax difference comparable to the Hyads discussed above. \citet{el-Badry2021} note that this criterion is also effective at rejecting unresolved multiples; this may be the case with the rejected pair TIC 25902832/25837464, as we observe two rotation periods in the TESS light curve for TIC 25902832. Our TESS sample ultimately consists of 11 cpm pairs (Table~\ref{tab:gaia-tess}).

Lastly, we search the TIC to determine if any of our targets have bright neighbors within 50" that may affect the observed light curve. We find that TIC 206327797 is separated by 29" from the late K-type star TIC 206327795, which shares its parallax and proper motion. The light curve centered on TIC 206327795 also shows the same periodic variability as the light curve centered on TIC 206327797, but the amplitude of the signal is a factor of 5 smaller; i.e., the variability is diluted if we choose an aperture that maximizes the light from TIC 206327795. We therefore attribute the periodicity to TIC 206327797. We do not see evidence of a second rotation period attributable to TIC 206327795.

TIC 43734215 is separated by 43" from the unassociated giant TIC 43734198, which has a comparable TESS magnitude. As the giant is not a short cadence target, we instead compare the Quick Look Pipeline (QLP; \citealt{Huang2020}) light curves centered at each of the two sources, extracted from the full-frame images (FFIs). While both light curves exhibit the observed periodicity, the amplitude is larger in the light curve centered on the M dwarf. We therefore attribute the periodicity to TIC 43734215.

The pair TIC 84731806/84731362 is located in a dense region, with both having nearby, unassociated giants fainter than themselves by roughly 0.9 TESS magnitudes, among other, fainter sources. TIC 84731806 is separated by 14" from giant TIC 84731756, while TIC 84731362 is separated by 31" from the giant TIC 84731326. Comparison of the QLP light curves for these sources is inconclusive. Instead, we use the \texttt{eleanor} package \citep{Feinstein2019} to create pixel-by-pixel light curves from the TESS target pixel files, from which we establish that the M dwarfs are the sources of the variability in both cases.

\subsection{Refining the MEarth sample}
The MEarth Project is a ground-based photometric monitoring survey of nearby M dwarfs, with telescope arrays in both the northern and southern hemispheres \citep{Nutzman2008, Irwin2015}. The team has identified rotation periods for 396 M dwarfs in the northern hemisphere and 251 in the south, finding periods between 0.1 and 140 days \citep{Newton2016, Newton2018, Medina2020, Medina2022}. While the MEarth target list consists of sources originally thought to be within 33pc, some of these stars are now known to be more distant after precise parallax measurements from \textit{Gaia}.

Some of our cpm pairs have rotation periods measured in \citet{Newton2016, Newton2018}, but additional data have also been collected since the publication of those works. We therefore crossmatch our cpm sample with MEarth and attempt to measure rotation periods for stars with recent observations using the methods described in \citet{Irwin2011}. In some cases, only one of the stars is a MEarth target; however, we are sometimes able to extract a usable light curve for the second object from the MEarth images. By combining the MEarth results with data in the MAST archive from TESS (and one system from K2), we extend our total sample to 25 M-M binary systems with measured rotation periods from either instrument; these are discussed individually in Section~\ref{sec:MM} and tabulated in Table~\ref{tab:main}. We note that 3 of our 11 pairs from the TESS sample also have MEarth data (TIC 114953216/114985772, TIC 256419669/52183206, and TIC 334637014/334637029) and will be referenced using their non-TIC identifiers in the analysis below. We also summarize the information available in the literature for each pair, with particular attention to H\textalpha\ and $v$sin$i$ measurements. We use the convention that a negative H\textalpha\ measurement indicates emission.

\begin{deluxetable*}{llrrrrr}[t]
\tabletypesize{\footnotesize}
\tablecolumns{7}
\tablewidth{0pt}
 \tablecaption{Masses and rotation periods for 25 nearby wide M-M multiples}
 \tablehead{
 \colhead{ \vspace{-0.1cm}Comp A} & 
 \colhead{ \vspace{-0.1cm}Comp B} &
 \colhead{$M_1$} &
 \colhead{$M_2$} &  
 \colhead{$P_{\rm{rot},1}$} & 
 \colhead{$P_{\rm{rot},2}$} &
 \colhead{$\rho$}
 \\
 \colhead{} &
 \colhead{} &
 \colhead{[M$_\odot$]} &
 \colhead{[M$_\odot$]} &
 \colhead{[days]} &
 \colhead{[days]} &
 \colhead{["]}}
\startdata
\multicolumn{6}{c}{\emph{Binaries}} \\
\\
TIC 22819180 & TIC 22819191 & 0.18 & 0.14 & 0.617 & 0.787 & 146 \\
TIC 197569385 & TIC 197570145 & 0.17 & 0.15 & 0.443 & 0.585 & 359 \\
TIC 37664980 & TIC 37664990 & 0.27 & 0.23 & 1.729 & 1.562 & 256 \\
TIC 206617113 & TIC 206617096 & 0.22 & 0.20 & 0.490 & 1.041 & 531 \\
TIC 43734215 & TIC 43789224 & 0.50 & 0.26 & 1.201 & 0.287 & 606 \\
TIC 84731806 & TIC 84731362 & 0.31 & 0.30 & 1.993 & 3.034 & 173 \\
TIC 450297524 & TIC 416857959 & 0.53 & 0.50 & 22 & 17 & 133 \\
LP 68-239 & 2MASS J15421300+6537051 & 0.43 & 0.42 & 2.207 & 0.617 & 233 \\
LHS 1377 & LHS 1376 & 0.40 & 0.27 & 11.019 & 3.023 & 106 \\
2MASS J21005492-4131438 & 2MASS J21010380-4114331 & 0.27 & 0.20 & 8.95 & 1.059 & 1036 \\
G 115-68 & G 115-69 & 0.21 & 0.20 & 0.748 & 0.854 & 7 \\
LP  167-64 & LP  167-63 & 0.41 & 0.14 & 57.66 & 0.995 & 47 \\
LP 613-49 & LP 613-50 & 0.36 & 0.12 & 11.76 & 1.66 & 25 \\
G 116-72 & G 116-73 & 0.36 & 0.33 & 0.755 & 0.974 & 23 \\
LP 12-72 & LP 12-90 & 0.53 & 0.16 & 1.050 & 1.044 & 960 \\
GJ 669 A & GJ 669 B & 0.44 & 0.28 & 20.51 & 1.457 & 17 \\
LHS 3808 & LHS 3809 & 0.33 & 0.14 & 94 & 1.569 & 12 \\
GJ 49 & GJ 51 & 0.59 & 0.20 & 19 & 1.024 & 294 \\
\\
\multicolumn{6}{c}{\emph{Confirmed \& candidate higher-order multiples}} \\
\\
LP 329-20 & LP 329-19 & 0.41 & $\leq$0.26 & 39.14 & 0.534/0.444 & 105 \\
GJ 810 AC & GJ 810 B & 0.20/0.25 & 0.14 & 137.37 & 134.63 & 107 \\
TIC 206327797 & TIC 206327810 & $\leq$0.62 & 0.14 & 0.224 & 1.523 & 308 \\
G 32-37 & G 32-38 AB & 0.34 & 0.30/0.17 & 34.01 & 1.592 & 16 \\
2MASS J07473239+4808438 & 2MASS J07473462+4807300 & $\leq$0.38 & 0.30 & 52.54 & 54.56 & 77 \\
2MASS J15483685-5045256 &  2MASS J15483762-5045143 & $\leq$0.42 & 0.17 & 61.19 & 103 & 13 \\
LP 90-130 & LP 90-129 & 0.23 & $\leq$0.22 & 8.046 & 0.518 & 35
\enddata
\label{tab:main}
\centering{
\tablecomments{See Section~\ref{sec:unc} for discussion of uncertainties.}}
\end{deluxetable*}

\vspace{-0.9cm}
\subsection{Rotation periods of M-M pairs}
\label{sec:MM}

\subsubsection{TIC 22819180 \& 22819191}
This pair of 0.18M$_\odot$ and 0.14M$_\odot$ stars is separated by 146" and located at a distance of 24pc. Both were observed in TESS sector 23. The light curve for TIC 22819180 exhibits a rotation period of 0.617 days, while TIC 22819191 shows a 0.787-day modulation.

TIC 22819180 was previously studied by HATNet, which reported a rotation period of 0.38 days \citep{Hartman2011}. We do not see this signal in the TESS light curve, although its value is consistent with the 1/(1+1/$P$) alias for a 0.617-day period. As TESS is space-based, it is not susceptible to 1-day aliases, unlike ground-based surveys like HATNet and MEarth.

\subsubsection{TIC 197569385 \& 197570145}

This pair has masses of 0.17M$_\odot$ \& 0.15M$_\odot$, separated by 359" and at 42pc. The stars were observed in TESS sector 28, where we measure a period of 0.443 days for TIC 197569385 and 0.585 days for TIC 197570145.

\needspace{6em}
\subsubsection{TIC 37664990 \& 37664980}
This pair has masses of 0.23M$_\odot$ \& 0.27M$_\odot$, separated by 256" and at a distance of 35pc. TIC 37664990 was observed in TESS sectors 3 \& 42, while TIC 37664980 was observed in sector 3 only. TIC 37664990 has a rotation period of 1.562 days while TIC 37664980 has a rotation period of 1.729 days.

\subsubsection{TIC 206617113 \& 206617096}
These stars have masses of 0.22M$_\odot$ \& 0.20M$_\odot$, a separation of 531", and a distance of 35pc. TIC 206617113 was observed in TESS sectors 2 \& 29, while TIC 206617096 was observed in sector 2 only. We measure a rotation period of 0.490 days for TIC 206617113 and a rotation period of 1.041 days for TIC 206617096.

\subsubsection{TIC 206327797 \& 206327810}
This pair with masses of 0.62M$_\odot$ \& 0.14M$_\odot$, distance 25pc, and separation 308" represents two components of a triple system. The primary, HIP 116491, is separated from TIC 206327797 by 29"; see discussion in Section~\ref{sec:tess}. TIC 206327810 was observed in TESS sector 29, from which we extract a rotation period of 1.523 days. TIC 206327797 was observed in TESS sectors 1 and 29. While the largest periodogram peak for TIC 206327797 corresponds to 0.112 days, a smaller peak at 0.224 suggests that the higher frequency signal is a strong second harmonic of the 0.224-day rotation period. The H\textalpha\ emission of TIC 206327797 was studied in \citet{Riaz2006}, who measured an equivalent width of -11.2\AA, consistent with its rapid rotation.

However, we caution that TIC 206327797 may itself be a binary, and therefore note this system under the `Confirmed \& candidate higher-order multiples' heading in Table~\ref{tab:main}. While it was not rejected with our RUWE $<2$ cut, it has RUWE = 1.7, the highest of any star remaining in the TESS sample. Furthermore, its \textit{Gaia} $B_P-R_P$ color is 2.77 mag, anomalously red for a 0.62M$_\odot$ M dwarf \citep[][updated web version \href{https://www.pas.rochester.edu/~emamajek/EEM_dwarf_UBVIJHK_colors_Teff.txt}{2021.03.02}]{Pecaut2013}, although overluminosity due to youth could also explain this discrepancy.

\subsubsection{TIC 43734215 \& 43789224}

This pair at 35pc is separated by 606", with component masses of 0.50M$_\odot$ \& 0.26M$_\odot$. Both stars were observed in TESS sector 26, from which we measure rotation periods of 1.201 days for TIC 43734215 and 0.287 days for TIC 43789224.

\citet{Bowler2019} measured H\textalpha\ emission with equivalent width -6.9\AA\ for TIC 43734215; this activity is consistent with the observed rapid rotation.

\subsubsection{TIC 84731806 \& 84731362}

This pair at 33pc has similar masses of 0.31M$_\odot$ \& 0.30M$_\odot$ and a separation of 173". TIC 84731806 was observed in TESS sector 14, from which we measure a rotation period of 1.993 days. TIC 84731362 was observed in sectors 14 \& 41; we measure a rotation period of 3.034 days.

\subsubsection{TIC 450297524 \& 416857959}
\label{sec:lightkurve}

This pair at 39pc has masses of 0.53M$_\odot$ \& 0.50M$_\odot$ and a separation of 133". Both were observed in TESS sector 21 and exhibit variability on the order of a TESS sector.

For TIC 416857959, we measure a candidate period of 16.5 days using the SAP light curve, although this signal is removed along with the TESS systematics in the PDCSAP light curve. To verify that this signal is true astrophysical variability, we remove the TESS systematics from the SAP light curve using the \texttt{Lightkurve} cbvCorrector (cotrending basis vector) functionality and the over- and under-fitting metrics provided in that package.\footnote{This method is described in the \href{http://docs.lightkurve.org/tutorials/2-creating-light-curves/2-3-how-to-use-cbvcorrector.html}{\texttt{Lightkurve} v2.0 tutorial}} We find that the candidate variability cannot be explained by TESS systematics without dramatically overfitting, with the \texttt{Lightkurve} method returning a period of 16.6 days. As we only observe one rotation period with TESS, we cannot say with certainty that this signal is not a harmonic of the true variation. However, this star was also previously studied by HATNet, which identified a rotation period of 17.1 days \citep{Hartman2011}. Including an MCMC in our \texttt{Lightkurve} fit, we estimate that the uncertainty in our measured rotation period is $\pm 1.4$ days. As the HATNet result is consistent with the TESS observations within uncertainties, we therefore assert that TIC 416857959 has a rotation period of roughly 17 days.

For TIC 450297524, the SAP light curve shows a periodicity of 10.5 days while the \texttt{Lightkurve} analysis yields 10.6 days. However, when we include an MCMC we find there is roughly equal posterior density in 11-day and 22-day peaks, suggesting that the 10.6-day signal may be a strong second harmonic. HATNet has also observed this star, measuring a rotation period of 22.5 days \citep{Hartman2011}. As the HATNet observations support the second harmonic hypothesis, we take the rotation period of TIC 450297524 to be roughly 22 days.

\citet{Ansdell2015} identified TIC 416857959 as young based on its NUV luminosity and measured marginal H\textalpha\ activity with an equivalent width of -0.83\AA.

\needspace{6em}
\subsubsection{2MASS J15421300+6537051 \& LP 68-239}

This pair at 39pc has similar masses of 0.42M$_\odot$ \& 0.43M$_\odot$ and a separation of 233". Both stars have been observed in 9 TESS sectors (14-17, 21-24, 41), with a rotation period of 0.617 days for 2MASS J15421300+6537051 and 2.207 days for LP 68-239.

While \citet{Newton2016} assigned 2MASS J15421300+6537051 a grade of `N' (non-detection or undetermined detection), they did identify 0.617 days as the most significant candidate period. They also measured the 2.207-day periodicity of LP 68-239 using 112 MEarth observations.

\citet{Newton2017} measured an H\textalpha\ equivalent width of -5.24\AA\ for LP 68-239, consistent with its rapid rotation. While 2MASS J15421300+6537051 was not studied in this work, it was also found to be active by \citet{Bowler2019}, who measured an H\textalpha\ equivalent width of \hbox{-7.3\AA}. \citet{Fouque2018} measured the $v$sin$i$ of 2MASS J15421300+6537051 to be 9.2 km/s, consistent with our measurement of the rotation period for an inclination $i$ of roughly 20$^{\circ}$.

\subsubsection{LHS 1376 \& LHS 1377}
LHS 1376 and LHS 1377 are a nearby pair at 13pc with masses of 0.27M$_\odot$ \& 0.40M$_\odot$, separated by 106". MEarth has collected 25,312 observations of this pair, from which we measure rotation periods of 3.023 days for LHS 1376 \& 11.019 days for LHS 1377. This system has not been observed by TESS.

We observed LHS 1376 four times with the CHIRON spectrograph ($R$=80000) on the 1.5m CTIO/SMARTS telescope as part of a spectroscopic survey of 0.1--0.3M$_\odot$ M dwarfs within 15pc \citep{Winters2021}, finding a $v$sin$i$ of 4km/s. This velocity is consistent with the measured rotation period for an inclination $i$ of roughly 60$^{\circ}$. We also measure a median H\textalpha\ equivalent width of -2.7\AA\ using the method of \citet{Medina2020}. While we did not observe LHS 1377, both stars were observed by \citet{Gaidos2014}, who found H\textalpha\ equivalent widths of -1.55\AA\ for LHS 1377 and -2.94\AA\ for LHS 1376.

\subsubsection{2MASS J21005492-4131438 \& 2MASS J21010380-4114331}

2MASS J21005492-4131438 \& 2MASS J21010380-4114331 are a widely-separated (1036") pair at 19pc with masses of 0.27M$_\odot$ \& 0.20M$_\odot$. The B component has a rotation period of 1.059 days in MEarth (4059 observations), which is also observed in TESS (sectors 1 and 27). MEarth only collected 87 observations of the A component and the stars are too widely separated for A to appear in the images of B, but A was observed in the two TESS sectors with a 8.95-day periodicity.

The H\textalpha\ emission of these stars was studied in \citet{Riaz2006}. They measured equivalent widths of -4.2\AA\ for A and -6.4\AA\ for B.

\subsubsection{G 115-68 \& G 115-69}

G 115-68 \& G 115-69 are a pair of stars with similar masses (0.21M$_\odot$ \& 0.20M$_\odot$) at 19pc, separated by 7". The stars were observed in TESS sector 21, albeit in the same TESS pixel. Two periodic signals appear in the TESS light curve, a stronger signal with $P=0.854$ days and a weaker signal with $P=0.748$ days. The stars were observed individually in MEarth (262 observations), revealing that the $P=0.854$-day signal is associated with G 115-69 and the $P=0.748$-day signal with G 115-68.

The H\textalpha\ activity for both stars were measured in \citet{Reid1995}: -4.19\AA\ for G 115-68 and -5.75\AA\ for G 115-69, consistent with their rapid rotation.

\subsubsection{LP 167-63 \& LP 167-64}
LP 167-63 \& LP 167-64 are 0.14M$_\odot$ \& 0.41M$_\odot$ M dwarfs at 24pc, separated by 47". Based on 1517 observations from MEarth, \citet{Newton2016} measured the rotation period of LP 167-64 to be 57.66 days but did not identify a statistically-significant rotation period in LP 167-63. LP 167-63 was observed in sector 21 of TESS, revealing a 0.995-day period. Such a signal is unlikely to be retrieved in MEarth given its proximity to the 1-day alias.

\citet{Reid1995} measured an H\textalpha\ equivalent width of -4.93\AA\ for LP 167-63, consistent with the rapid rotation we measure. \citet{Newton2017} found LP 167-64 to be inactive in H\textalpha\ (EW $= 0.131$\AA), consistent with its slow rotation.

\subsubsection{LP 613-49 \& LP 613-50}

LP 613-49 and LP 613-50 are a pair at 19pc with masses of 0.36M$_\odot$ \& 0.12M$_\odot$ and a separation of 25". Both components were observed in K2, with rotation periods measured in \citet{Armstrong2015}. The authors find a rotation period of 11.76 days for LP 613-49 and 1.66 days for LP 613-50. LP 613-49 was also observed 675 times by MEarth, with \citet{Newton2016} identifying a 11.66-day rotation period, consistent with the K2 result. LP 613-50 was not a MEarth target, although it was observed in the images of LP 613-49. While we attempted to extract a rotation period from these images, we were unable to identify a statistically-significant signal. This is unsurprising, as the exposure times were optimized for the brighter LP 613-49. The stars are blended in TESS sector 46, from which we measure an 11.7 day rotation period using the \texttt{Lightkurve} method discussed in Section~\ref{sec:lightkurve}. We are unable to retrieve a second rotation period from the blended light curve.

Both components were identified as H\textalpha-active in \citet{Reid1995}, with the authors measuring an equivalent width of -2.56\AA\ for LP 613-49 and -2.22\AA\ for LP 613-50.

\needspace{6em}
\subsubsection{G 116-72 \& G 116-73}

G 116-72 \& G 116-73 have masses of 0.36M$_\odot$ \& 0.33M$_\odot$, separated by 23". The pair is located at 27pc. \citet{Newton2016} identify a rotation period of 0.755 days for G 116-72 using 1464 observations from MEarth. While G 116-73 is not a MEarth target, it was observed in the full-frame images of G 116-72. As the pair is similar in mass, G 116-73 is sufficiently bright in the G 116-72 images that we are able to extract a 0.974-day rotation period. We detect both rotation periods in the blended light curve from TESS sector 21.

A $v$sin$i$ of 22.5km/s was measured for G 116-72 by \citet{Kesseli2018}. This velocity is consistent with the measured rotation period for an inclination $i$ of roughly 70$^{\circ}$. \citet{Hawley1996} measure an H\textalpha\ equivalent width of -15.49\AA\ for G 116-72 and -6.84\AA\ for G 116-73.

\needspace{6em}
\subsubsection{LP 12-72 \& LP 12-90}
LP 12-72 \& LP 12-90 (0.53M$_\odot$ \& 0.16M$_\odot$) are part of a hierarchical triple system, along with HD 220140, a G-type star. The two M dwarfs are separated by 960", while LP 12-72 is separated by 10" from the G star \citep{Makarov2007}. The system is at 19pc. The \texttt{BANYAN \textSigma} kinematics tool \citep{Gagne2018} suggests that this system may be part of the Columba young association, with an age of 42 Myr \citep{Bell2015}. Given the overluminosity expected for such young stars, the masses we quote from the \citet{Benedict2016} relation would overestimate the true masses.

Both M dwarfs were observed in \citet{Newton2016} using MEarth (903 observations), with the authors finding a rotation period of 1.050 days for LP 12-72 and a rotation period of 1.044 days for LP 12-90. This rotation period for LP 12-90 is also detected in four TESS sectors (19, 24, 25, 26). LP 12-72 is contaminated by the G-star primary in TESS; for the G star, we measure a rotation period of 2.7 days.

The lithium abundance of the G star was studied in \citet{Xing2021}, who measured an equivalent width of 198m\AA. Using the \texttt{BAFFLES} code \citep{StanfordMoore2020} to convert this abundance to an age posterior, we estimate an age of 80 Myr. There is also power in the posterior at younger ages, consistent with membership in the 42-Myr Columba association.

\citet{Alonso-Floriano2015} observed high H\textalpha\ activity levels in both M dwarfs, measuring equivalent widths of -7.5\AA\ for LP 12-72 and -11.2\AA\ for LP 12-90. A $v$sin$i$ of 25.4km/s was observed for LP 12-72 in \citet{Jeffers2018}, which is consistent with our observed rotation period assuming a roughly edge-on inclination.

\subsubsection{G 32-37 \& G 32-38}

G 32-37 and G 32-38 are a pair of M dwarfs at 29pc, separated by 16". Both were observed by MEarth as part of \citet{Newton2016}, who measured rotation periods of 34.01 days and 1.592 days, respectively, using 863 MEarth observations.

However, G 32-38 is itself a binary, with two components separated by 0.151" revealed by lucky imaging \citep{Janson2014}. Deblending the components prior to using the \citet{Benedict2016} $K$-band relation, these components have masses of 0.30M$_\odot$ and 0.17M$_\odot$, while G 32-37 has a mass of 0.34M$_\odot$.
 
We are able to forward model the significant peaks in the MEarth periodogram of G 32-38 using a sinusoidal light curve with $P=1.592$, a second sinusoid with $P/3$, and the window function. This strong third harmonic is also seen in the TESS light curve (sectors 17, 42, 43). We do not identify a second candidate rotation period. It is unclear whether the 0.17M$_\odot$ or the 0.30M$_\odot$ component is responsible for the observed 1.592-day periodicity.

\citet{Newton2017} investigated the H\textalpha\ activity of \hbox{G 32-37} and G 32-38 AB, measuring an equivalent width of -0.740\AA\ for G 32-37 (marginally active, consistent with its intermediate $P_{\rm rot}$) and -7.404\AA\ for \hbox{G 32-38 AB} (very active, consistent with rapid rotation).

\subsubsection{GJ 669A \& GJ 669B}
GJ 669 A \& B are a nearby pair at 11pc with masses of 0.44M$_\odot$ and 0.28M$_\odot$. They are separated by 17". The two stars were observed with MEarth in \citet{Newton2016}, yielding a rotation period of 20.51 days for A and 1.457 days for B from 1045 observations. Both a 1.457-day periodicity and a 20-day sinusoidal trend are detected in the blended TESS light curve (sectors 25 and 26), although the latter is too long to be considered robustly-detected by TESS (i.e., the length of the period is similar to the length of a TESS sector).

We obtained 10 observations of GJ 669 A and 9 observations of GJ 669 B using the TRES spectrograph ($R=44000$) on the FLWO 1.5m telescope as part of a spectroscopic survey of nearby, mid/low-mass M dwarfs \citep{Winters2021}. We do not resolve any rotational broadening of GJ 669 A, which is consistent with its 20.51-day rotation period at the resolution of the spectrograph. We measure a $v$sin$i$ of 7km/s for GJ 669 B; this velocity is consistent with the measured $P_{\rm rot}$ for an inclination $i$ of roughly 40$^{\circ}$. Both stars have H\textalpha\ activity; we measure a modest equivalent width of -1.93\AA\ for A and stronger activity with EW$= -7.69$\AA\ for B.

\subsubsection{LHS 3808 \& LHS 3809}
This pair at 23pc has masses 0.33M$_\odot$ \& 0.14M$_\odot$ and a separation of 12". \citet{Newton2016} identified a 1.569-day period for LHS 3809 using 444 MEarth observations, but LHS 3808 was given a `U' grade indicating that the candidate 90.14-day rotation period was uncertain. To date, 8680 MEarth observations have been collected of this pair. The additional observations continue to support the candidate periodicity from \citet{Newton2016}; we therefore adopt our refined estimate of 94 days for LHS 3808 and recover the previously-reported 1.569-day periodicity for LHS 3809.

Both components were observed in a blended light curve in TESS sector 42. We identify 1.575 days as the most significant period in this blended light curve, in agreement with the MEarth value for LHS 3809.

LHS 3809 was identified as highly H\textalpha-active in \citet{Newton2017}, with the authors measuring an equivalent width of  -15.506\AA. While this work did not measure an H\textalpha\ equivalent width for LHS 3808, a previous study did not detect any H\textalpha\ emission for this source \citep{Reid1995}. These activity levels are consistent with expectations given that the observed rotation periods indicate LHS 3808 has spun down and LHS 3809 has not.

\subsubsection{LP 329-19 \& LP 329-20}
This pair at 28pc is separated by 105", with component masses of 0.26M$_\odot$ \& 0.41M$_\odot$ for LP 329-19 \& LP 329-20, respectively. Both were observed in \citet{Newton2016}, who identified rotation periods of 39.14 days for LP 329-20 and 0.534 days for LP 329-19 using 1575 MEarth observations.

However, LP 329-19 has a RUWE of 3.2, indicating it is likely an unresolved binary (although the separation between the components cannot be very small given that there is an astrometric perturbation, and so we are not concerned with binary interactions). Some of the $\Delta m-$separation parameter space is ruled out by a null detection in the Robo-AO survey \citep{Lamman2020}. This star was observed in TESS sectors 24 and 25; four prominent peaks appear in the TESS Lomb Scargle periodogram: 0.534 days and its second harmonic, and 0.444 days and its second harmonic. These peaks also appears in the MEarth periodogram. It therefore appears that both close components are rapid rotators, and both are less massive than LP 329-20 regardless of the unknown light ratio. In the limit of an equal mass binary, both components would have masses of 0.17M$_\odot$, although an equal-mass binary would not produce an astrometric perturbation.

The H\textalpha\ activity of this system was studied in \citet{Newton2017}. They found that LP 329-20 was inactive, measuring an H\textalpha\ equivalent width of 0.049\AA. One or both of the LP 329-19 components is active, with an equivalent width of -4.707\AA\ in the blended spectrum. This result is consistent with the observed rotation rates.

\needspace{6em}
\subsubsection{2MASS J07473462+4807300 \& 2MASS J07473239+4808438}

This pair at 29pc has masses 0.30M$_\odot$ \& 0.38M$_\odot$, separated by 77". Both were investigated in \citet{Newton2016} using 101 MEarth observations, but given `U'-grade classifications that indicate that their estimated rotation periods were uncertain. In particular, the authors identified a tentative 54.56-day period for the less massive component, 2MASS J07473462+4807300, and 1.017 days for 2MASS J07473239+4808438. To date, 779 MEarth observations have been collected. The classification for 2MASS J07473462+4807300 appears robust. For 2MASS J07473239+4808438, we favor $P_{\rm rot}$= 52.54 days.

Both stars were observed in TESS sector 20. We do not identify any periodic signals for either star in the TESS data, and in particular, do not observe a 1.017-day periodicity for 2MASS J07473239+4808438. The TESS light curves are therefore consistent with our assertion that both stars are slow rotators. The candidate 1.017-day period reported by \citet{Newton2016} is likely the 1/(1-1/$P$) alias.

While neither star has a published close companion, we note that 2MASS J07473239+4808438 has a \textit{Gaia} RUWE of 4.62. It is therefore likely that this component is an unresolved binary. In the limit of an equal-mass binary, each component would have mass 0.25M$_\odot$; there is therefore the possibility that 2MASS J07473462+4807300 is the most massive star in the system. Some of the $\Delta m-$separation parameter space is ruled out by a null detection in the Robo-AO survey \citep{Lamman2020}. 

The \texttt{BANYAN \textSigma} tool finds a high probability that this system is a member of the Argus association with an age of 40--50 Myr \citep{Zuckerman2019}. Such a young age would be surprising given the slow rotation we measure. This calculation assumes the radial velocity of the system is not known, and so Bayesian inference is used to marginalize over this parameter to compute membership probabilities. If the system is truly a member of Argus, the tool predicts a radial velocity of around 15km/s. If one adopts the 5.7km/s radial velocity from APOGEE \citep{Birky2020}, the probability of Argus membership drops to near zero. We conclude that this pair is not a member of Argus.

\needspace{6em}
\subsubsection{GJ 810A \& GJ 810B}

GJ 810 A \& GJ 810 B are a nearby pair at 12pc with a separation of 107". Both were observed with MEarth in \citet{Newton2018}, who found a rotation period of 137.37 days for A and 134.63 days for B using 5676 MEarth observations. GJ 810 A is itself a double-lined spectroscopic binary \citep{Baroch2018} with a separation of 0.0916" \citep{Vrijmoet2022}. Deblending the components yields masses of 0.25M$_\odot$ \& 0.20M$_\odot$ for A and C, while B is the least massive star in the system with a mass of 0.14M$_\odot$. We cannot definitively establish whether the observed rotation period is associated with component A or C.

We have obtained 4 observations of GJ 810 AC and 4 observations of GJ 810 B using the TRES spectrograph ($R=44000$) as part of a spectroscopic survey of nearby, mid/low-mass M dwarfs \citep{Winters2021}. We do not see H\textalpha\ emission in the composite AC spectra, suggesting that A and C are likely both inactive slow rotators. The B component is also H\textalpha-inactive, with a median equivalent width of 0.0\AA\ and no observable rotational broadening. The AC spectra also appear consistent with no rotational broadening, although the $v$sin$i$ measurement is challenging due to the blended lines.

\subsubsection{2MASS J15483685-5045256 \& 2MASS J15483762-5045143}

2MASS J15483685-5045256 and 2MASS J15483762-5045143 are a pair at 46pc with masses of 0.42M$_\odot$ \& 0.17M$_\odot$ and a separation of 13". 2MASS J15483685-5045256 was observed with MEarth in \citet{Newton2018}, who found a rotation period of 61.19 days using 1922 MEarth observations. We extract a light curve of 2MASS J15483762-5045143 from the MEarth images, from which we identify a candidate period of 103 days for the lower-mass component. While there is no published third component in this system, 2MASS J15483685-5045256 has a \textit{Gaia} EDR3 RUWE of 2.49, suggesting that this component is likely an unresolved binary.

\needspace{6em}
\subsubsection{GJ 49 \& GJ 51}

GJ 49 \& 51 are a pair of 0.59M$_\odot$ \& 0.20M$_\odot$ stars at 10pc, separated by 294". \citet{Newton2016} identified a 1.024-day rotation period for GJ 51 using 434 MEarth observations. We measure the same value using TESS sectors 18 and 24. While GJ 49 was also investigated in \citet{Newton2016}, it was given a `U' grade indicating that the candidate 0.738-day rotation period was uncertain. The TESS sectors do not show a short-period signal, instead exhibiting modulation with a 9.45-day period. However, we note the PDCSAP TESS light curves can have the low-frequency stellar variability removed along with systematics; while the SAP light curves may be contaminated with systematics, they retain low-frequency stellar variability. The SAP light curve suggests that the 9.45-day periodicity may be a harmonic of the signal, and the actual rotation period is 18.9 days. With the benefit of 2520 MEarth observations collected in 2021, we identify a 19.0-day modulation, although the peak in the periodogram is broad. We also observe peaks associated with the aliases 1/(1+1/$P$) and 1/(1-1/$P$). A rotation period of 18.4 $\pm$ 0.7 days was measured by \citet{Mascareno2018} using ASAS. Taken together, there is consistent evidence that GJ 49 has a rotation period of roughly 19 days.

We have collected 11 observations of GJ 51 with the TRES spectrograph as part of our spectroscopic survey of nearby, mid/low-mass M dwarfs \citep{Winters2021}. We measure a $v$sin$i$ of roughly 11km/s, consistent with the measured rotation period for an inclination $i$ around 80$^\circ$. We measure a median H\textalpha\ equivalent width of \hbox{-11.11\AA,} consistent with its rapid rotation.

\citet{Reiners2018} did not observe rotational broadening in their spectrum of GJ 49, consistent with a rotation period greater than 14 days. \citet{Gizis2002} report that GJ 49 is inactive in H\textalpha, measuring an equivalent width of 0.283\AA.

\subsubsection{LP 90-129 \& LP 90-130}
This pair with similar masses of 0.22M$_\odot$ \& 0.23M$_\odot$ is located at 32pc, with a separation of 35" between components. LP 90-129 was studied in \citet{Newton2016}, who identified a rotation period of 0.518 days using 1562 MEarth observations. While LP 90-130 was not a MEarth target, it was captured in the images of LP 90-129, from which we extract a rotation period of 8.046 days. The pair was also observed in TESS sectors 20 and 21, from which we identify both an 8.2-day and 0.518-day periodicity.

LP 90-129 has a \textit{Gaia} EDR3 RUWE of 2.9, suggesting that this component is likely a binary; \citet{Newton2016} also flagged this star as potentially overluminous. In the limit of an equal-mass binary, both components would have masses of 0.15M$_\odot$.

\citet{Newton2017} found at least one of the LP 90-129 components to be active, measuring an H\textalpha\ equivalent width of -4.461\AA. This activity is consistent with the observed rapid rotation.

\subsubsection{Spin-orbit coupled binaries}

We exclude systems with known close companions from our list of M-M pairs, as interactions between close binary components can modify rotation rates. In this section, we briefly discuss two systems with close binaries and remark on the implications for our sample.

GJ 1006 A \& B are a pair with separation 25" at 15pc while GJ 1230 A \& B are separated by 5" and located at 10pc. In both cases, the A component is a close binary: from spectroscopic observations, \citet{Baroch2018} measured an orbital period of $3.956523^{+0.000071}_{-0.000092}$ days for GJ 1006 AC while \citet{Delfosse1999} measured an orbital period of 5.06880±0.00005 days for GJ 1230 AC. Using the mass ratios reported in these works and the iterative deblending technique described in \citet{Winters2019, Winters2021}, we estimate masses of 0.34M$_\odot$ \& 0.12M$_\odot$ for GJ 1006 A \& C and 0.26M$_\odot$ \& 0.24M$_\odot$ for GJ 1230 A \& C. The \citet{Benedict2016} relation yields masses of 0.28M$_\odot$ for GJ 1006 B and 0.20M$_\odot$ for GJ 1230 B.

GJ 1230 A \& C are spin-orbit coupled. From 7678 MEarth observations, we measure a rotation period of 5.027 days. The pair (along with the B component, due to their small angular separation) was also observed in TESS sector 40; this light curve exhibits a 5.083-day periodicity. Considering uncertainties, these measurements are equivalent to the 5.069-day orbital period. We do not observe a second (or third) candidate period in the TESS light curve. While GJ 1230 B was observed in a separate MEarth light curve, we are unable to identify a statistically-significant periodicity. However, we have collected four observations of GJ 1230 B with the TRES spectrograph as part of a spectroscopic survey of 0.1--0.3M$_\odot$ M dwarfs within 15pc \citep{Winters2021}. We find marginal H\textalpha\ emission with a median equivalent width of -0.71\AA\ and no detectable rotational broadening at the resolution of the spectrograph, suggesting the GJ 1230 B is likely spun down. Meanwhile, we observe two H\textalpha\ peaks in the blended GJ 1230 AC spectra, indicating that both stars are active.

GJ 1006 AC was observed in \citet{Newton2016}, who identified a rotation period of 4.798 days based on 946 MEarth observations. A similar rotation period of 4.7901 days was identified by SuperWASP \citep{Norton2007}. The system has not been observed by TESS. This rotation rate is marginally slower than the 4-day orbital period. At least one of the pair must be H\textalpha\ active, with \citet{Shkolnik2009} measuring an H\textalpha\ equivalent width of -2.7\AA\ in the composite spectrum. Meanwhile, \citet{Newton2016} identified 96.610 days as a candidate rotation period for GJ 1006 B, with a `U'-grade classification. With the benefit of additional MEarth data set collected since the publication of that work, we confirm a rotation period of roughly 93.5 days. \citet{Newton2017} found GJ 1006 B to be H\textalpha-inactive, with an equivalent width of 0.089\AA.

\begin{deluxetable*}{llrrrlrrrr}[t]
\tabletypesize{\footnotesize}
\tablecolumns{10}
\tablewidth{0pt}
 \tablecaption{Properties of the white dwarf -- M dwarf pairs \label{tab:wd}}
 \tablehead{
 \colhead{ \vspace{-0.1cm}M} & 
 \colhead{$M_{\rm M}$} &
 \colhead{$P_{\rm rot}$} &
 \colhead{H\textalpha\ EW} &
 \colhead{$\rho$} &
 \colhead{ \vspace{-0.1cm}WD} & 
 \colhead{$t_{\rm cool}$} &
 \colhead{$t_{\rm C18}$} &
 \colhead{$t_{\rm min}$} &
 \colhead{$M_{\rm WD}$}
 \\
 \colhead{} &
 \colhead{[M$_\odot$]} &
 \colhead{[days]} &
 \colhead{[\AA]} &
 \colhead{["]} &
 \colhead{} &
 \colhead{[Gyr]} &
 \colhead{[Gyr]} &
 \colhead{[Gyr]} &
 \colhead{[M$_\odot$]}}
\startdata
SCR J1107-3420 B & 0.26 & 7.611$^{a}$ & & 31 & SCR J1107-3420 A & 0.32 & 1.5 & 0.82 & 0.64 \\
LHS 2928 & 0.25$^*$ & 126.233$^{a}$ & & 37 & LHS 2927 & 5.4 & 16.9 & 5.9 & 0.51 \\
2MASS J23095781+5506472 & 0.14 & 104.1$^{b}$ & -0.56$^{d}$ & 6 & 2MASS J23095848+5506491 & 4.2 & 5.3 & 4.7 & 0.65 \\
G 68-34 & 0.46 & 0.655$^c$ & -5.05$^{e}$ & 9 & LP 463-28 & 5.0 & 6.7 & 5.5 & 0.60 \\
GJ 1179 B** & 0.12 & & -0.01$^f$ & 188 & GJ 1179 A & 4.9 & 103 & 5.4 & 0.43 \\
GJ 283 B & 0.10 & & -0.14$^g$ & 20 & GJ 283 A & 1.4 & 2.1 & 1.9 & 0.71 \\
GJ 754.1 B & 0.26 & & 0.14$^f$ & 27 & GJ 754.1 A & 0.62 & 1.2 & 1.1 & 0.73 \\
GJ 166 C & 0.24 & & -4.42$^f$ & 8 & GJ 166 B & 0.13 & 3.6 & 0.63 & 0.56 \\
GJ 169.1 B** & 0.31 & & 0.08$^f$ & 10 & GJ 169.1 A & 2.3 & 3.1 & 2.8 & 0.72  
\enddata
\tablecomments{$t_{\rm C18}$ is the estimated total age using the \citet{Cummings2018} initial-final mass relation and may be inaccurate in our mass regime (and is clearly unphysical in some cases, e.g. GJ 1179 A). $t_{\rm min}$ is the minimum age of the system based on the cooling age of the white dwarf and the expectation that the initial mass must be $<3$M$_\odot$ to produce the observed white dwarf mass.\newline
$^{*}$ This star has a RUWE of 5.5, suggesting a blended component that may mean the mass is overestimated \newline
$^{**}$ As the white dwarf is the primary, we list the M dwarf as the B component; note that these designations are reversed on Simbad \newline
$^{a}$\citealt{Newton2018},
$^{b}$\citealt{Newton2016},
$^{c}$Measured from 297 MEarth observations,
$^{d}$\citealt{Newton2017},
$^{e}$\citealt{Reid1995},
$^{f}$Measured with TRES using method from \citet{Medina2020},
$^{g}$Measured with CHIRON using method from \citet{Medina2020}}
\end{deluxetable*}

\vspace{-0.9cm}

In the context of this work, these systems highlight the difficulties posed by unresolved close binaries. If the C components of these systems were not known, we might be surprised that the B components are spun down, inactive, and presumably old, yet the more massive A components remain active and rapidly rotating. In truth, it is likely that close binary interactions are causing the activity and rapid rotation to persist.

Furthermore, these systems show that rotation rate and activity are intrinsically linked, as opposed to both independently depending on age; i.e., an old star whose rapid rotation is maintained due to spin-orbit coupling also maintains its H\textalpha\ activity.

\subsection{A note on uncertainties}
\label{sec:unc}

Our mass uncertainties are dominated by the uncertainty in the \citet{Benedict2016} MLR, which has an RMS error of 0.014M$_\odot$ calculated over the entire relation (0.08--0.62M$_\odot$). In the region of highest scatter (near 0.2M$_\odot$), the authors find errors of $\pm$0.035M$_\odot$, which they attribute to differences in age, composition, and magnetism within their calibration sample.

\citet{Newton2016, Newton2018} do not report uncertainties for their MEarth rotation periods, noting that there are typically multiple peaks in the periodogram and therefore an estimate of the uncertainty based on the width of the most significant peak would be misleading. Simulations of MEarth data in \citet{Irwin2011} suggest typical uncertainties of $< 1$\% for $P_{\rm rot} < 10$ days, $1$\% for $10 < P_{\rm rot} < 20$ days, 2\% for $20 < P_{\rm rot} < 50$ days, 5--10\% for $50 < P_{\rm rot} < 100$ days, and 20--30\% for $P_{\rm rot} > 100$ days. These ranges are therefore representative of the uncertainties in our MEarth rotation periods, although they may be conservative given improvements to MEarth since 2011.

We can only measure relatively short rotation periods with TESS given the 27-day duration of each sector. To estimate representative uncertainties for these periods, we include an MCMC in our \texttt{Lightkurve} fits and measure the 68\% interval of the posteriors, with the caveat stated above that this method can underestimate the true uncertainty; particularly for the longer rotation periods, there is a risk that we are misidentifying a harmonic as the fundamental mode. For $P_{\rm rot} < 1$ day, we find typical uncertainties of $<0.1$\%, increasing to a few tenths of a percent for periods of a few days, to 1--5\% for periods up to half a TESS sector. Periods longer than half a TESS sector are very uncertain and have the highest risk of being a mistaken harmonic.

If a star has its rotation measured by both TESS and MEarth, the TESS period is likely more accurate for periods less than a few days and the MEarth period is likely more accurate for periods longer than a few days. We populate Table~\ref{tab:main} based on this principle.

\section{WD-M binaries}
\label{sec:wd}
While the binaries discussed thus far allow us to investigate $P_{\rm rot}$ as a function of mass for coeval field stars, the age of those stars is unknown. Next, we consider a complementary sample: M dwarfs with earlier-type primaries, for which we can estimate the age of the primary star, and hence the age of the M dwarf. In this section, those earlier-type primaries are white dwarfs; in Section~\ref{sec:fgk-intro}, they are FGK stars.

We identify nine WD-M cpm pairs with MEarth observations of the M dwarf (Table~\ref{tab:wd}). Of these nine M dwarfs, we are able to detect a rotation period for four. While these four systems are the focus of this section, one of the M dwarfs without a measured rotation period also merits some discussion. This star, GJ 166 C, is active in H\textalpha\ \citep{Gizis2002} and highlights a potential shortcoming of aging M dwarfs based on wide white-dwarf companions. The WD-M pair is part of a widely-separated triple, with the third component being a K-type star. Based on the chemistry of the K star, \citet{Fuhrmann2014} argue that the system is old and the present activity of GJ 166 C is the result of mass/angular momentum transfer from the white dwarf when it was formed, despite the large separation between these components (a projected separation of 42 AU). If this is the case, we are making an unjustified assumption by asserting that the rotation period of the widely-separated M dwarf is unaffected by its primary; however, we note that this white dwarf, GJ 166 B, has the shortest cooling age and the smallest projected physical separation from its M-dwarf companion out of the nine white dwarfs we investigated, characteristics that may make GJ 166 C most susceptible to this feedback.

The other four M dwarfs without measured rotation periods are inactive in H\textalpha\ and therefore likely slow rotators. While we will not discuss them further in this section, they are included in Table~\ref{tab:wd} as their properties may be of general interest.

For each white dwarf, we use \textit{Gaia} EDR3 colors/magnitudes and the \texttt{WD\_models} Python package\footnote{\href{https://github.com/SihaoCheng/WD_models}{https://github.com/SihaoCheng/WD\_models}} to estimate cooling ages and white dwarf masses using the \citet{Bedard2020} cooling tracks. This package then calculates the total age using the \citet{Cummings2018} initial-final mass relation (IFMR) and the MIST models \citep{Choi2016}. However, the \citet{Cummings2018} IFMR is monotonic; recent work has indicated that there may be a kink \citep{Marigo2020} or large dispersion \citep{Barrientos2021} at the low masses relevant for our sample ($M_f \leq 0.73$M$_\odot$, corresponding to $M_i \leq 3$M$_\odot$). As a result, the main sequence lifetimes---and hence, the total ages---of our white dwarfs are uncertain. For example, a 0.65M$_\odot$ white dwarf could be produced by a 1.6, 2.1, or 2.4 M$_\odot$ main-sequence star using the non-monotonic \citet{Marigo2020} IFMR. The \citet{Cummings2018} IFMR estimates an initial mass of 2.0M$_\odot$ for this star.

While the total age is subject to this uncertainty, the cooling age provides a robust limit. Furthermore, a main-sequence star more massive than 3M$_\odot$ is in the monotonic regime of the IFMR and cannot produce the white dwarf masses we measure. We therefore add 0.5 Gyr (roughly, the main sequence lifetime of a 3M$_\odot$ star) to the cooling ages to obtain lower limits on the \hbox{M-dwarf age.}

The conversion between the position of a white dwarf on the HR diagram and its cooling age is dependent on composition. We follow the method of \citet{Fleury2022} and ascribe a composition (H-dominated, He-dominated, or mixed) based upon which model produces the minimal $\chi^2$ in \citet{GentileFusillo2021}, who fit these atmospheric models to \textit{Gaia} astrometry and photometry. Using this criterion, we classify LP 463-28 and GJ 169.1 A as He-dominated and the others
as H-dominated. However, LP 463-28 was analyzed spectroscopically by \citet{Limoges2013}, who classified this white dwarf as spectral type DA; we therefore override its classification to be H-dominated. GJ 169.1 A
was spectroscopically identified as a He white dwarf in \citet{Limoges2015}. These classifications ultimately have a marginal impact on our results; switching compositions yields differences in cooling ages of a few percent for most of our targets, with differences as high as thirty percent for stars with very old cooling ages.

\begin{deluxetable*}{lrrrlrlrrlrrrr}[t]
\tabletypesize{\footnotesize}
\tablecolumns{14}
\tablewidth{0pt}
 \tablecaption{Properties of the active M dwarfs with FGK primaries}
 \tablehead{
 \colhead{\vspace{-0.07cm}M} &
 \colhead{$M$} &
 \colhead{H\textalpha} &
 \colhead{$P_{\rm rot}$} &
 \colhead{\vspace{-0.07cm}Ref} &
 \colhead{$\rho$} &
 \colhead{\vspace{-0.07cm}FGK} &
 \colhead{$B-V$} & 
 \colhead{$P_{\rm rot}$} &
 \colhead{ \vspace{-0.07cm}Ref} &
 \colhead{Li EW} &
 \colhead{$T_{\rm eff}$} &
 \colhead{log$g$} &
 \colhead{[M/H]}
 \\
 \colhead{} &
 \colhead{[M$_\odot$]} &
  \colhead{[\AA]} &
 \colhead{[days]} &
 \colhead{} &
 \colhead{["]} &
 \colhead{} &
 \colhead{[mag]} &
 \colhead{[days]} &
 \colhead{} &
 \colhead{[m\AA]} &
 \colhead{[K]} &
 \colhead{[dex]} &
 \colhead{[dex]}}
\startdata
G 271-110 & 0.31 & -7.67 & 1.06 & 1 & 612\phantom{$^*$} & HD 10008 & 0.797 & 7.1 & 1,6 & 94.5 $\pm$ 1.6 & 5392 & 4.61 & 0.13 \\
LP 99-392 & 0.31 & -5.92 & 1.196 & 2 & 43\phantom{$^*$} & HD 139477 & 1.063 & 11.2 & 1 & 5.6 $\pm$ 2.0 & 4835 & 4.72 & -0.06 \\ 
LP 128-32 & 0.32 & -5.44 & 6.5 & 1,2 & 178\phantom{$^*$} & HD 93811 & 0.937 & 8.5 & 1,6 & 20.5 $\pm$ 2.4 & 5098 & 4.55 & -0.12 \\
GJ 166 C & 0.24 & -4.42 &  &  & 78\phantom{$^*$} & GJ 166 A & 0.820 & 42 & 7 & $<$ 1.1 & 5191 & 4.61 & -0.28 \\
LP 876-10 & 0.20 & -4.06 & 0.318 & 1,3 & 26738* & HD 216803 & 1.094 & 10.1 & 1,6 & 29.8 $\pm$ 1.7 & 4739 & 4.68 & -0.10 \\
LSPM J0849+0329W & 0.23 & -4.08 & 4.4 & 1 & 159\phantom{$^*$} & HD 75302 & 0.689 & 16.4 & 8 & 16.8 $\pm$ 2.5 & 5735 & 4.54 & -0.01 \\
G 232-62 & 0.27 & -2.25 & 59 & 4 & 77\phantom{$^*$} & HD 211472 & 0.810 & 8.5 & 1 & 19.8 $\pm$ 2.4 & 5366 & 4.62 & -0.10 \\
2MASS J22562702+7600101 & 0.19 & -2.89 & 2.51 & 1 & 242\phantom{$^*$} & HD 217142 & 0.942 & 20.3 & 1 & $<$ 4.0 & 5090 & 4.63 & -0.10 \\
2MASS J05363846+1117487 & 0.29 & -3.81 & 5.3 & 5 & 157\phantom{$^*$} & HD 245409 & 1.415 & 11.2 & 1 & 34.9 $\pm$ 2.6 & 3966 & 4.70 & 0.05 \\
HD 183870 B & 0.24 & -3.17 & 2.32 & 2 & 209\phantom{$^*$} & HD 183870 & 0.922 &  &  & $<$ 4.1 & 5107 & 4.65 & -0.06 
\enddata
 \label{tab:fgk}
\tablecomments{Rotation references:
$^{1}$Measured from TESS,
$^{2}$Measured from MEarth,
$^{3}$\citet{Newton2018},
$^{4}$Measured from ZTF,
$^{5}$Measured from TESS FFI,
$^{6}$\citet{Strassmeier2000},
$^{7}$\citet{Rosenthal2021},
$^{8}$\citet{Wright2011}
\newline$^*$ While this separation is very large, \citet{Mamajek2013} identify this system as a true multiple with an interloper probability of 10$^{-5}$}
\end{deluxetable*}

\begin{figure*}[t]
\vspace{-0.5cm}
    \centering
    \includegraphics[width=\textwidth]{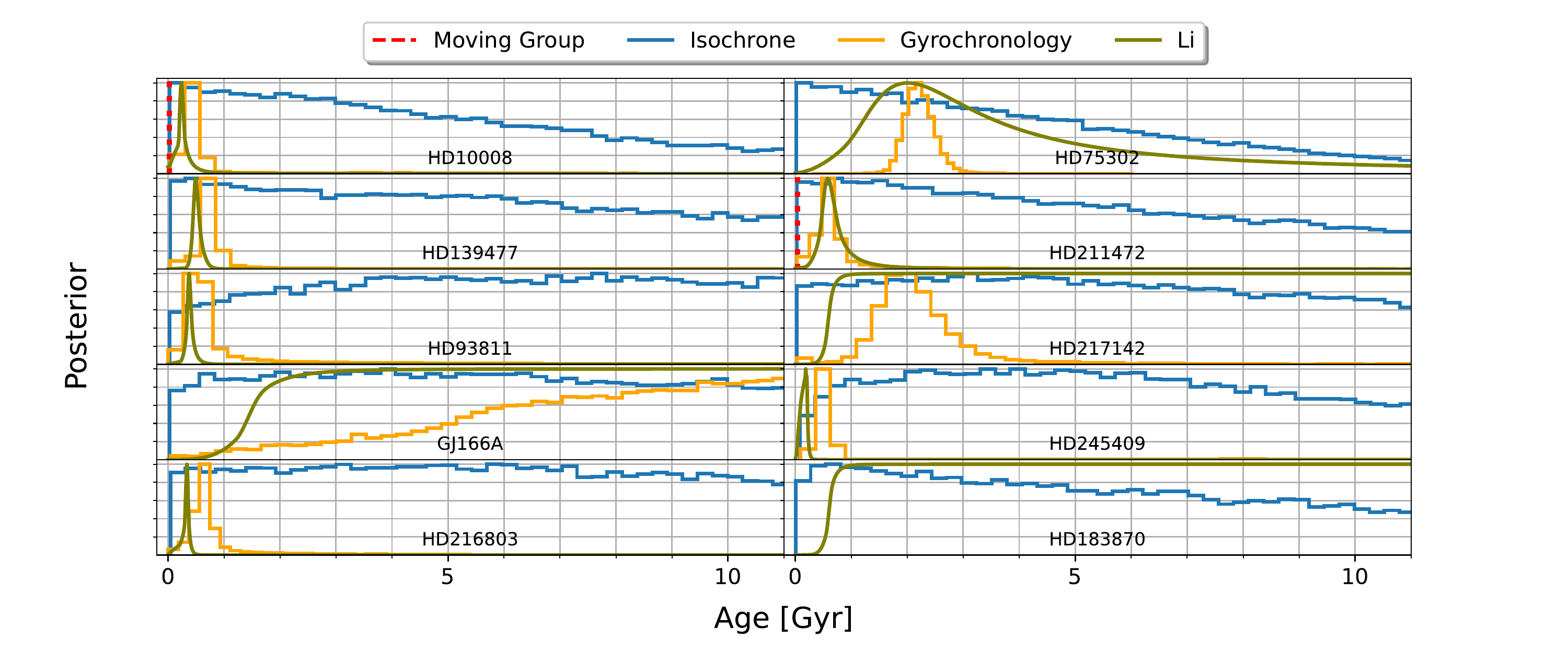}
    \caption{Age posteriors for wide, FGK primaries to active, low-mass M dwarfs, based on isochrones and gyrochronology from \texttt{stardate}, lithium abundance aging from \texttt{BAFFLES}, and moving group membership from \texttt{BANYAN \textSigma}. Six of the ten stars have posteriors that are consistent with young ($\leq$ 600 Myr) ages, similar to the ages probed by studies of M dwarfs in clusters. Four (GJ 166 A, HD 75302, HD 217142, and HD 183870) are consistent with ages older than 1 Gyr.}
    \vspace{0.2cm}
    \label{fig:ages}
\end{figure*}

\vspace{-0.85cm}

The two M dwarfs with long-period determinations (LHS 2928 and 2MASS J23095781+5506472) are both old, with minimum ages of 5.9 Gyr and 4.7 Gyr, respectively. SCR J1107-3420 B exhibits an intermediate rotation period of 7.611 days, with a minimum age of 820 Myr. The \citet{Cummings2018} IFMR results in an age of 1.5 Gyr, although this estimate should be interpreted loosely given the large uncertainties in the IFMR at low masses. These three systems are consistent with a model of spindown in which low-mass M dwarfs transition quickly from rapidly to slowly rotating at an age of a few Gyr.

G 68-34 is curious: this 0.46M$_\odot$ M dwarf exhibits rapid rotation with a period of 0.655 days, yet the system has a minimum age of 5.5 Gyr. The white dwarf is old regardless of assumptions on composition, with the adoption of a He-dominated model only reducing the cooling age by 0.5 Gyr. The activity and rapid rotation of this M dwarf cannot be explained using the mechanism proposed by \citet{Fuhrmann2014} for \hbox{GJ 166 C}; even if the formation of the white dwarf resulted in spinup of the M dwarf, that spinup would have occurred $t_{\rm cool} = 5.0$ Gyr ago. It is unlikely that we have misidentified the rotation period in the MEarth data, as the H\textalpha\ activity measured by \citet{Reid1995} is consistent with rapid rotation. It is possible that the M dwarf is a binary; \citet{Khovrichev2018} identify the star as a candidate binary using SDSS data, with a separation of 1.8" and a $\Delta m$ of 1.132 mag in $u$-band. If true, this would correspond to masses of roughly 0.38M$_\odot$ and 0.24M$_\odot$. However, \textit{Gaia} should have been able to image both components at such a separation, and so the lack of a detection in EDR3 makes this hypothesis unlikely; also, such a companion would not explain the observed rapid rotation. \textit{Gaia} does not rule out the possibility of a binary companion at a smaller separation; while this star does not have a large RUWE, a very close companion would not induce an astrometric perturbation.

\section{FGK-M Binaries}
\label{sec:fgk-intro}
As part of an ongoing effort to measure the metallicities of fully-convective M dwarfs, we had previously gathered $R=44000$ TRES spectra of pairs of stars consisting of an M dwarf secondary and an FGK primary. Of these M dwarfs, ten are active based on their H\textalpha\ emission (measured using the method described in \citealt{Medina2020}); we therefore anticipate that these ten systems are young, but we can further quantify this youth by studying the more massive primaries.

For nine of the ten primaries, we estimate the effective temperature, metallicity, and surface gravity from their TRES spectra using the Stellar Parameter Classification (SPC) tool \citep{Buchhave2012}, with uncertainties of 0.1 dex in log$g$, 50K in temperature, and 0.08 dex in [M/H]. The tenth (HD 245409), a late K dwarf, is cooler than the minimum temperature for which SPC is calibrated. We therefore adopt the stellar properties from the \citet{Mann2015} analysis of this star, which is based on medium-resolution optical and NIR spectra. We then use these spectral properties and photometry from \textit{Gaia} EDR3 to estimate isochronal ages using the \texttt{stardate} package \citep{Angus2019}, although we find that these estimates are ultimately uninformative. We also measure the rotation period of the primary from TESS, or adopt a rotation period from the literature for all primaries but HD 183870, which has not yet been observed by TESS and which lacks a literature measurement (Table~\ref{tab:fgk}). We then use \texttt{stardate} to estimate gyrochronological ages (Figure~\ref{fig:ages}). We note that HD 183870 has an activity index of log$(R'_{HK})= -4.47$ \citep{Marsden2014}, which suggests an age of roughly 1 Gyr when coupled with its $B-V$ color and the gyrochromochrones from Figure 11 of \citet{Mamajek2008}.

\begin{figure*}
    \hspace{-1cm}
    \centering
    \includegraphics[width=\textwidth]{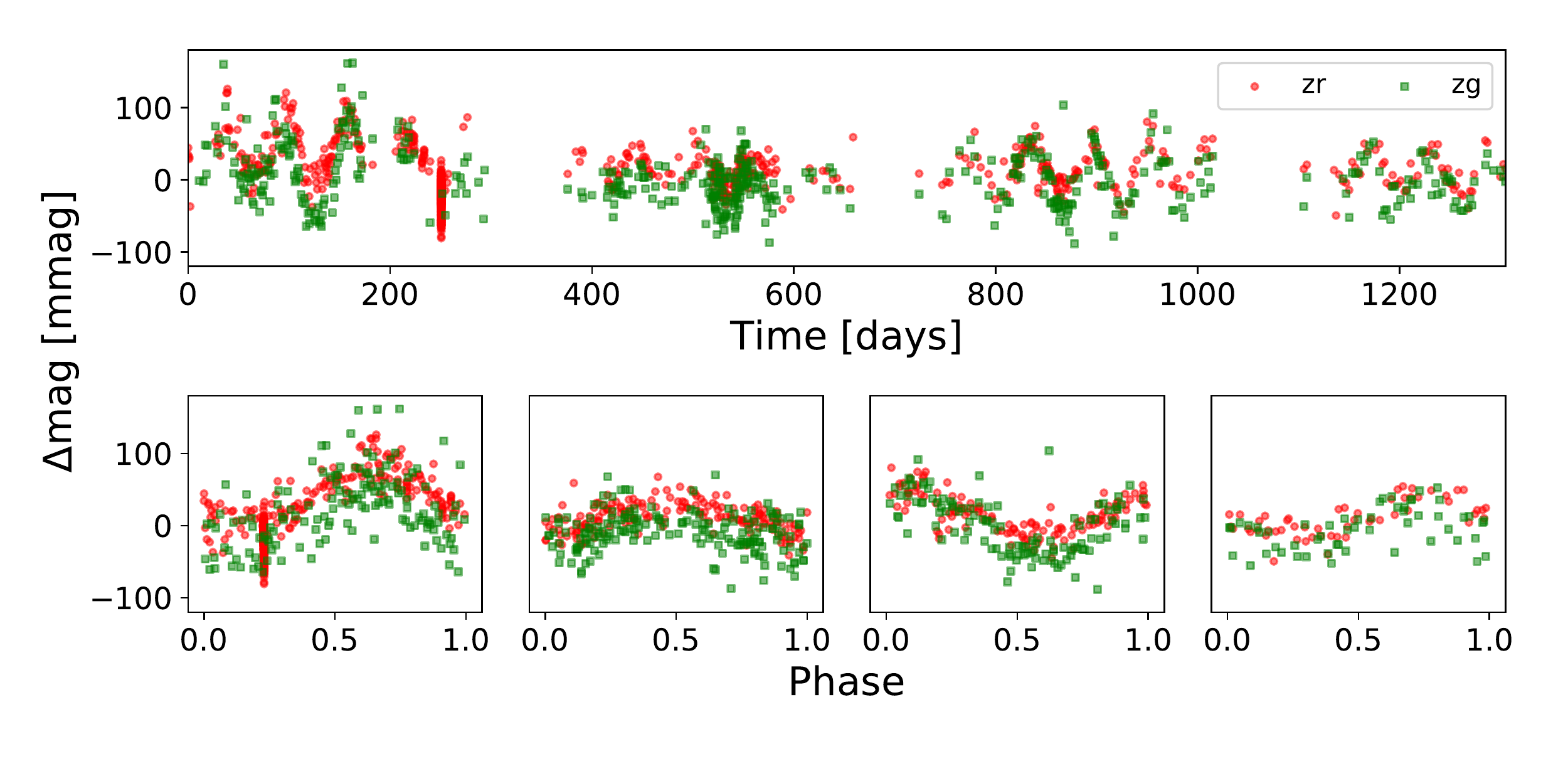}
    \vspace{-0.8cm}
    \caption{Rotational modulation of G 232-62, observed in two filters with the Zwicky Transient Facility \citep{Masci2019}. In the top panel, we show the photometry publicly available through ZTF DR9. In the bottom panel, we phase the data to a 59-day period. Each year is presented separately for clarity, as the spot pattern evolves over the four years of data.}
    \label{fig:ztf}
\end{figure*}

We also determine the rotation periods for most of the secondaries using analogous methods to those described in Section~\ref{sec:MM}; however, the analysis for G 232-62 is unique. This star was not observed with TESS at 2-minute cadence and we do not retrieve a rotation period from the full-frame TESS images. While the star was observed with MEarth in \citet{Newton2016}, only 225 observations were collected and the authors were not able to establish a conclusive rotation period. However, the star was monitored for multiple years by the Zwicky Transient Facility \citep[ZTF;][]{Masci2019}, from which we identify a 59-day rotation period; this rotational modulation is evident in both survey filters (Figure~\ref{fig:ztf}).

Based on gyrochronology of the primaries, all but \hbox{GJ 166 A} are younger than a few Gyr; as discussed in Section~\ref{sec:wd}, the activity of GJ 166 C may be the result of the recent evolution of the B component into a white dwarf, and not a marker of youth. However, the gyrochronological ages should be interpreted with caution. At young ages, \texttt{stardate} inflates uncertainties rather than modeling the range of periods at a given age, and it does not account for stalling at older ages \citep[e.g.,][]{Curtis2020}. That said, replacing the \texttt{stardate} estimates with those from the gyrochronological relation of \citet{Spada2020} yields similar results for most of our stars, with the exception that the latter relation produces a younger estimate for HD 245409.

We have two further reservations about the gyrochronological estimates: firstly, the longer rotation periods are taken from a variety of sources that may not be equally reliable (and in particular, the rotation period of HD 217142 is measured using TESS but is longer than half a TESS sector, a regime where we are susceptible to misidentifying a harmonic as the rotation period). Secondly, we use \texttt{BANYAN \textSigma} \citep{Gagne2018} to identify that two of the pairs are likely members of young moving groups, with HD 10008 \& G 271-110 members of the \textbeta\ Pic moving group with an age of 24$\pm$3 Myr and HD 211472 \& \hbox{G 232-62} members of the Argus association with an age of 40--50 Myr \citep{Zuckerman2019}; the probability of association membership is 92\% for HD 10008 and 96\% for HD 211472. The gyrochronological ages produced by \texttt{stardate} for these stars are an order of magnitude older than the age of their presumed host cluster.

To increase our confidence in our age assessments, we consider a second, independent age estimator: lithium absorption. We measure lithium equivalent widths from the TRES spectra of the FGK primaries, following the method described in \citet{Zhou2021}; that is, fitting Gaussians to the Li lines at 6707.76\AA\ and 6707.91\AA\ and the Fe I line at 6707.43\AA\ and using the area ratios to correct for the iron contamination of the lithium doublet (S.\ Quinn, private communication). We detect lithium absorption at $>2\sigma$ significance in seven of the ten stars and place upper limits for the remaining three. We then use the \texttt{BAFFLES} framework \citep{StanfordMoore2020} to convert these lithium abundances (or upper limits) and the star's $B-V$ color to an age posterior; we adopt the $B-V$ colors tabulated in the \textit{Hipparcos} or Tycho-2 catalogs \citep{vanLeeuwen2007, Hog2000}. We find that the lithium ages tend to be younger than the ages obtained from gyrochronology, although the methods are broadly in agreement as to which systems are younger and which are older. Given the possible biases in the gyrochronological relation at both young and older ages, we generally consider the lithium-based estimates to be more reliable. Interestingly, we note that both lithium and gyrochronology suggest an age of roughly 600 Myr for HD 211472, making it unlikely that this system is a member of the 40--50 Myr Argus Association. While the lithium observations of HD 10008 favor an age of roughly 240 Myr, there is some power in the posterior at very young ages, and so its membership in the 24-Myr \textbeta\ Pic moving group remains possible.

We also have a sample of 21 inactive M dwarfs with widely-separated FGK primaries from the same metallicity project. As we do not have rotation periods for these M dwarfs, they are generally uninformative for this analysis; however, there is one pair in this sample that is worthy of discussion. 2MASS J02580617+2040016 shows H\textalpha\ in absorption, with an equivalent width of 0.05\AA\ in our TRES spectra. However, \texttt{BANYAN \textSigma} indicates that it and its F-type primary, HD 18404, are both members of the Hyades, with an age of 600--800 Myr depending on the source (see comparison of literature estimates in \citealt{Douglas2019}). Unfortunately, we do not have a $P_{\rm rot}$ estimate for this star; 2MASS J02580617+2040016 was not observed by MEarth, and the TESS photometry is dominated by the F star.

We follow the treatment of Hyads in \citet{Douglas2016} and linearly interpolate between the $M_K$ and $M_*$ points given in \citet{Kraus2007}, obtaining a mass of 0.22M$_\odot$ for 2MASS J02580617+2040016. Alternatively, the \citet{Benedict2016} relation yields 0.29M$_\odot$. We discuss this discrepancy in more detail in Section~\ref{sec:pra}, ultimately finding that the \citet{Benedict2016} estimate is more accurate. We therefore adopt 0.29M$_\odot$ in the analysis below, although our conclusions do not change if the 0.22M$_\odot$ estimate is used instead.

\citet{Douglas2016, Douglas2019} found nearly all observed Hyads with $M_* < 0.3$M$_\odot$ were rapidly rotating, with the slowest rotators of this mass still possessing rotation periods $P_{\rm rot} < 30$ days. However, this result is inconsistent with the observed H\textalpha\ inactivity of 2MASS J02580617+2040016; \citet{Newton2017} found that a typical inactive star with mass 0.29M$_\odot$ would rotate with a period of 77 days, with a 22-day standard deviation in this trend. Some inactive stars with $M_* < 0.3$M$_\odot$ rotated with periods as short as 40 days, although this is still substantially longer than the periods observed in \citet{Douglas2016, Douglas2019}.

There are a few possible explanations for this discrepancy: i) 2MASS J02580617+2040016 is not a Hyad and simply shares its UVW space motion by coincidence; ii) 2MASS J02580617+2040016 is inactive but rapidly rotating; or, iii) some 0.29M$_\odot$ M dwarfs have spun down to rotation periods of $\geq 40$ days by 600 Myr, but they were not identified in works such as \citet{Douglas2016} due to the reduced sensitivity of missions like K2 and TESS to long rotation periods.

The first option would be surprising; there is consensus in the literature that the F-type primary, HD 18404, is a bona fide Hyades member based on its kinematics and metallicity \citep[e.g.,][]{Perryman1998, Gagne2018}. Furthermore, \texttt{stardate} returns gyrochronological and isochronal ages consistent with the Hyades age for this star. We do not detect lithium in the primary, but its $B-V$ color places it within the lithium dip at the age of the Hyades \citep{StanfordMoore2020} and so a non-detection is plausibly consistent. The second explanation would also be unusual, given that no rapidly-rotating, low-mass stars were identified in \citet{Newton2017} as having H\textalpha\ in absorption. We note that \citet{Newton2017} did report two stars with $M_* < 0.3$M$_\odot$, $P_{\rm rot} < 30$ days as inactive, but both these outliers can be explained. While 2MASS J22250174+3540079 was reported to have a mass of 0.28M$_\odot$, with the benefit of a precise parallax from \textit{Gaia}, that mass should be revised upwards to 0.43M$_\odot$. The other, LP 48-485, has been observed with TRES and we identify H\textalpha\ in emission, with an equivalent width of \hbox{-0.84\AA}. The -1\AA\ cutoff for activity used by \citet{Newton2017} is relatively arbitrary, with \citet{West2015} advocating for an inactive/active threshold of -0.75\AA. With the benefit of higher-resolution spectra where we can fully resolve the H\textalpha\ line, it is clear that LP 48-485 should be considered marginally active, unlike 2MASS J02580617+2040016. The fastest-rotating, inactive, $M_* < 0.3$M$_\odot$ star that remains in \citet{Newton2017} is LP 546-25, with a rotation period of 38 days. This star may be an analog for 2MASS J02580617+2040016.

We conclude that the most likely interpretation is that 2MASS J02580617+2040016 is slowly rotating, possibly with a period around 40 days, if not longer. This hypothesis should be tested with long-term photometric monitoring by an instrument that can resolve it from the bright primary at 15" separation.

\section{Discussion}
\label{sec:discussion}
\subsection{Comparison with M dwarfs in Praesepe}
\label{sec:pra}

\begin{figure}
    \centering
    \includegraphics[width=\columnwidth]{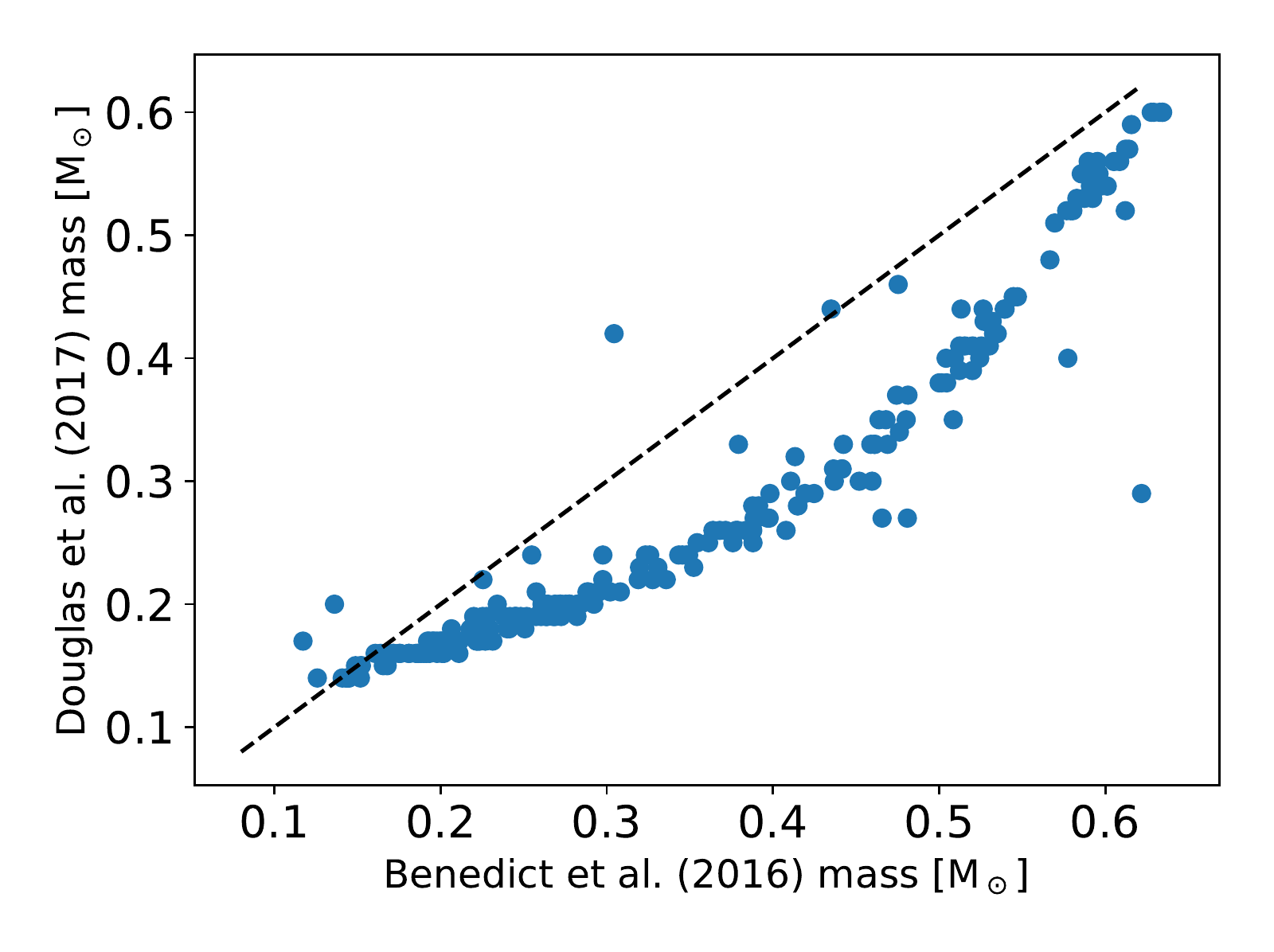}
    \vspace{-0.8cm}
    \caption{Comparison between the masses published in \citet{Douglas2017} for presumed single Praesepe M dwarfs using the \citet{Kraus2007} MLR and masses we estimate using the \citet{Benedict2016} MLR. The dashed line indicates unity. The outliers are the result of inaccurate distances for a few of the \citet{Douglas2017} estimates, as they adopted the cluster distance for targets where a \textit{Hipparcos} parallax was unavailable. Our estimates benefit from precise \textit{Gaia} parallaxes.}
    \label{fig:masses}
\end{figure}

Praesepe is a cluster with an age similar to the Hyades (600--800 Myr, depending on the source), with rotation periods for many of its constituent M dwarfs studied in \citet{Douglas2014, Douglas2017, Douglas2019}. As one of the oldest clusters with a well-studied M dwarf population (and with more M dwarfs with measured rotation periods than the Hyades), it provides a natural point of comparison to our field M dwarf sample. However, we need consistent mass estimates in order to compare these two samples. In the previous section, we noted that the mass of 2MASS J02580617+2040016 varied by 0.07M$_\odot$ depending on whether we used the \citet{Benedict2016} MLR or the \citet{Kraus2007} MLR adopted by  \citet{Douglas2014, Douglas2017, Douglas2019}. 2MASS J02580617+2040016 is not an exception; we compare the masses published in \citet{Douglas2017} for all the single Praesepe M dwarfs with our mass estimates from the \citet{Benedict2016} MLR, finding that the discrepancy is as large as 0.13M$_\odot$ for some masses (Figure~\ref{fig:masses}). This inconsistency is perhaps unsurprising given that the relations are calibrated using very different methods: the \citet{Benedict2016} MLR is calibrated using dynamical masses of astrometric binaries, while the \citet{Kraus2007} relation is based on stellar models.

\begin{figure}[h]
    \centering
    \includegraphics[width=\columnwidth]{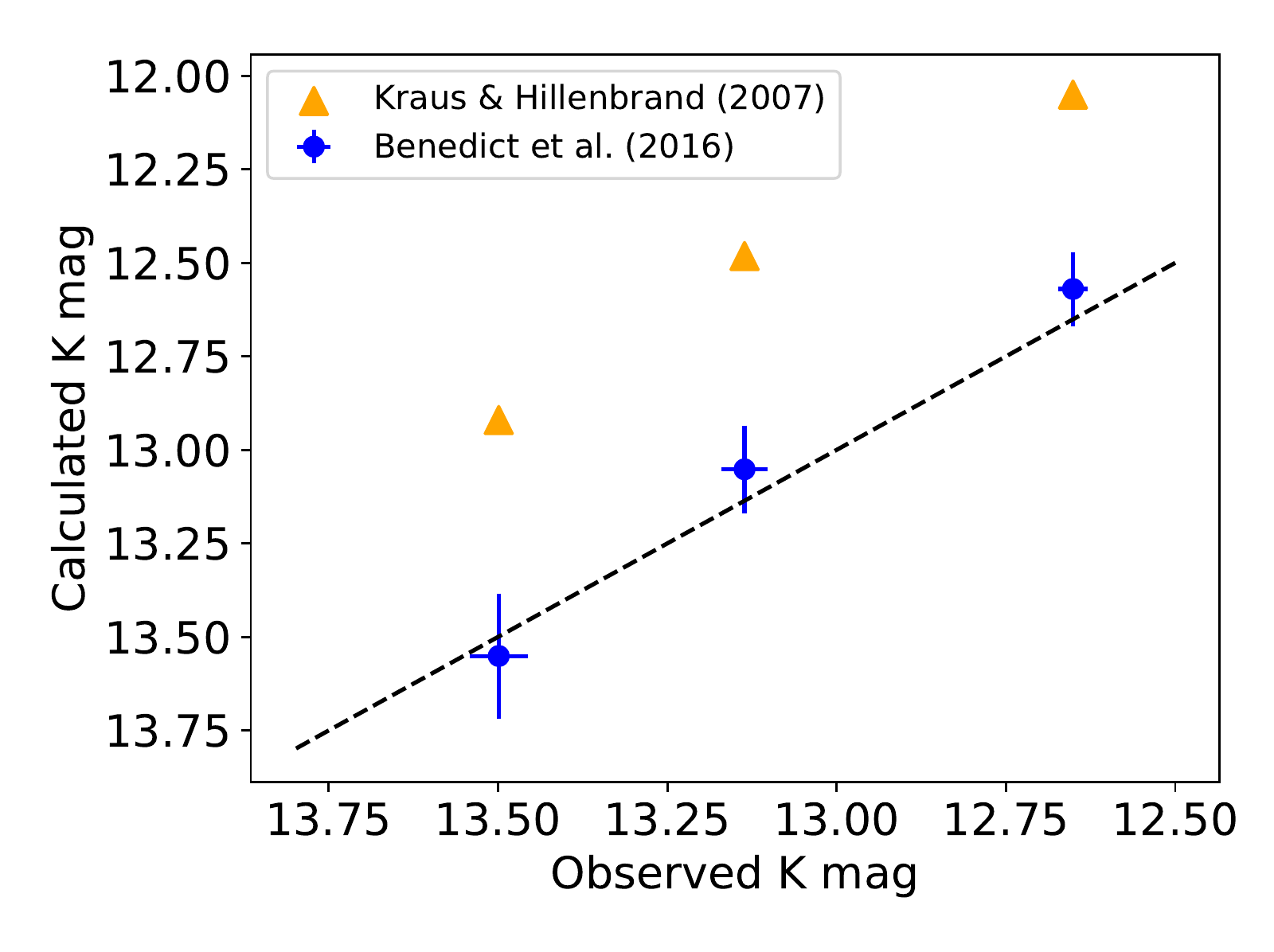}
    \vspace{-0.8cm}
    \caption{Observed and theoretical $K$-band magnitudes for the blended light from three Praesepe eclipsing binaries. The dashed line indicates unity. We consider the MLRs of \citet{Benedict2016} and \citet{Kraus2007}, finding good agreement between \citet{Benedict2016} and observations from 2MASS. The \citet{Kraus2007} relation predicts the systems would be brighter than observed by roughly 0.5 mag. For the \citet{Benedict2016} points, we propagate uncertainties in mass, parallax, and 2MASS magnitude, and add in quadrature the reported 0.09 mag RMS in the \citet{Benedict2016} MLR.}
    \label{fig:ebs}
\end{figure}

As Praesepe is somewhat young, it may be inappropriate to use a relation such as \citet{Benedict2016}, which is calibrated using main-sequence stars. Pre-main sequence stars are overluminous, which could cause the \citet{Benedict2016} relation to produce inflated mass estimates. However, this effect cannot explain Figure~\ref{fig:masses}; if overluminosity due to youth was creating the bias, we would expect the greatest discrepancy to occur at the lowest masses, with the relations falling into agreement at intermediate and high masses, where the M dwarfs are on the main sequence at the age of Praesepe.

\begin{figure*}[t]
    \centering
    \includegraphics[width=\textwidth]{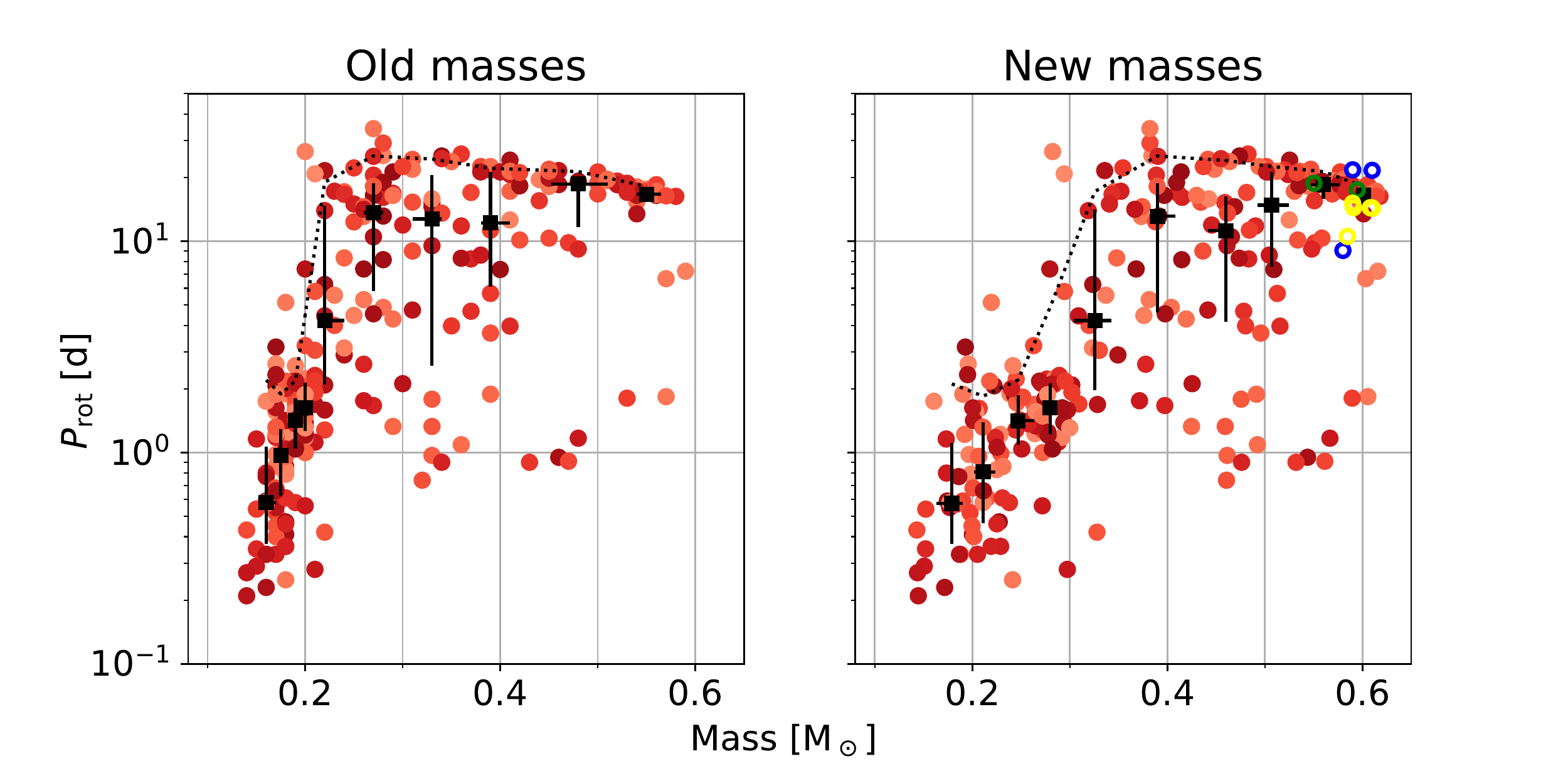}
    \vspace{-0.5cm}
    \caption{Masses and rotation periods for presumed single M dwarfs in the 600-Myr Praesepe cluster, one of the oldest clusters with measured photometric rotation periods for a large number of M dwarfs. Rotation periods are taken from \citet{Douglas2019}. In the left panel, masses are also taken from that work, while in the right panel, masses are estimated using the \citet{Benedict2016} $K$-band relation. Based on Praesepe eclipsing binaries, we argue that the masses in the right figure are correct (Figure~\ref{fig:ebs}). We use various shades of red such that individual points can be more easily distinguished; the same star is the same shade in both panels. Black rectangles denote median properties of bins containing equal numbers of stars, with error bars noting the interquartile range of masses and rotation periods in that bin. A dotted black line marks the 90th percentile of rotation periods in each bin; for stars with masses $>0.35$M$_\odot$, the most slowly-rotating stars appear to converge onto a slowly-rotating sequence. The open circles in the right panel show the small number of early M dwarfs in studies of older clusters from the literature, with masses again calculated using the \citet{Benedict2016} $K$-band relation: yellow denotes the 1.0 Gyr cluster NGC 6811 \citep{Curtis2019}; green denotes the 1.4 Gyr cluster NGC 752 \citep{Agueros2018}; blue denotes the 2.7 Gyr cluster Ruprecht 147 \citep{Curtis2020}. Those authors interpret the small differences between the rotation periods in Praesepe and in these older clusters as evidence of stalling in K dwarfs and early M dwarfs. We note that the 1.0 Gyr points fall below the slowly-rotating Praesepe sequence; this may suggest that properties other than age (e.g., metallicity) influence the slowly-rotating sequence for a cluster.}
    \label{fig:main}
\end{figure*}

To test the suitability of the \citet{Benedict2016} relation on Praesepe M dwarfs, we consider three of the Praesepe eclipsing binary systems published in \citet{Gillen2017}. We neglect the fourth system in that work, as the large uncertainties in the component masses make the system uninformative for our purposes. The systems we consider have components with the following precise dynamical masses: 0.3813$\pm$0.0074M$_\odot$ \& 0.2022$\pm$0.0045M$_\odot$; 0.212$\pm$0.012M$_\odot$ \& 0.255$\pm$0.013M$_\odot$; and 0.276$\pm$0.020M$_\odot$ \& a brown dwarf, which we assume to be negligible in its effect on the system luminosity. We convert each component to an absolute $K$-band luminosity using the \citet{Benedict2016} MLR, use the \textit{Gaia} EDR3 parallax to convert absolute to apparent magnitude, and determine the expected magnitude of the blended source. We compare this expectation to the observed $K$-band magnitude from 2MASS \citep{Cutri2003}. We also perform this same procedure for the \citet{Kraus2007} relation, linearly interpolating between their published values of absolute $K$ magnitude versus mass. Figure~\ref{fig:ebs} shows the results of this analysis. We find that the \citet{Benedict2016} relation produces estimates that closely align with the observed 2MASS magnitudes, while the \citet{Kraus2007} relation predicts that the systems should be brighter than observed in the $K$ band by roughly 0.5 mag. Based on this analysis, we conclude that the \hbox{\citet{Benedict2016}} MLR is appropriate for use in Praesepe, at least for M dwarfs more massive than 0.2M$_\odot$. Overluminosity may still bias the mass estimates for the lowest-mass M dwarfs at the Praesepe age, but we are unable to test this with the \citet{Gillen2017} sample. Nevertheless, this analysis clearly favors the mass estimates from \citet{Benedict2016} over those of \citet{Kraus2007}.

We therefore adopt the rotation periods of Praesepe M dwarfs from \citet{Douglas2019}, but replace the mass estimates with those from \citet{Benedict2016}. We cut binaries using the flags in \citet{Douglas2019} and remove an additional four stars using a conservative limit on possible astrometric perturbations (RUWE $< 1.4$). Both the original and our updated mass-rotation diagrams are shown in Figure~\ref{fig:main}. We note that this mass revision has important implications for the spindown epoch of low-mass M dwarfs. \citet{Douglas2017} found that it was common for 0.2--0.3M$_\odot$ stars to have already begun spinning down at the age of Praesepe, which was perhaps surprising given the number of rapidly-rotating 0.2--0.3M$_\odot$ M dwarfs observed in field surveys such as \citet{Newton2017}. With the revised mass estimates, we find that very few of Praesepe's 0.2--0.3M$_\odot$ M dwarfs are rotating with $P_{\rm rot} > 2$ days.

From this mass-period diagram, we identify the feature referred to in \citet{Popinchalk2021} as the `elbow'---that is, the characteristic mass that separates the slowly-rotating sequence from the rapidly-rotating reservoir. For Praesepe, this elbow roughly coincides with the radiative-convective boundary at mass 0.35M$_\odot$. Works such as \citet{Irwin2007} and \cite{Popinchalk2021} have found that this elbow occurs at higher masses in younger clusters. It may be a coincidence that the elbow aligns with the fully-convective boundary at the age of Praesepe, or it could be a consequence of a change in spindown behavior for fully-convective stars; as the elbow has not been studied in older clusters, it is unknown whether it stalls at this mass. We also observe the reservoir of fast rotators at masses lower than the elbow and with rotation periods typically less than 2 days. Rotation period is not independent of mass in this reservoir; as highlighted by the binned data in Figure~\ref{fig:main}, there is a correlation of increasing rotation period with increasing mass for rapid rotators. The median 0.3M$_\odot$ star in Praesepe rotates with $P_{\rm rot} = 2$ days while the median 0.2M$_\odot$ star rotates with $P_{\rm rot} = 0.7$ days.

The correlation between mass and rotation period is imprinted on low-mass M dwarfs at early ages; \citet{Somers2017} identified a similar mass-rotation correlation for the low-mass M dwarfs in the \hbox{10-Myr} Upper Sco and the \hbox{125-Myr} Pleiades, which they interpret as resulting from the mass-independence of the stars' specific angular momentum at 1--2 Myr. This mass-rotation correlation persists through the spinup of the \hbox{10-Myr} M dwarfs to the rotation periods seen at \hbox{125 Myr} (as these stars contract onto the main sequence), and through the spindown to the periods observed in Praesepe.

\begin{figure*}
    \hspace{-1cm}
    \centering
    \includegraphics[width=0.9\textwidth]{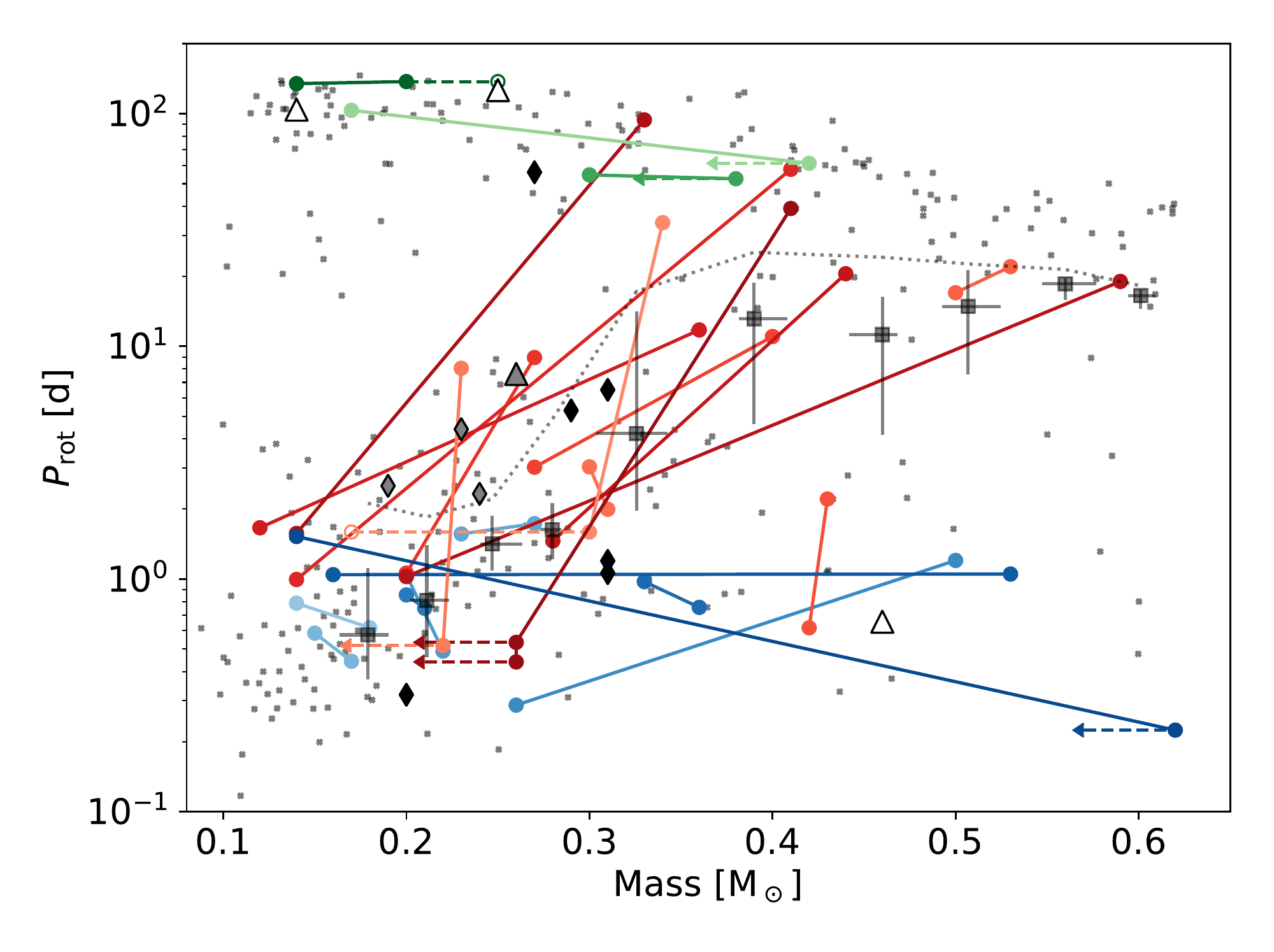}
    \vspace{-0.5cm}
    \caption{The masses and rotation periods for stars in our wide binary sample. The four triangles denote M dwarfs with a white dwarf companion from Table~\ref{tab:wd}, the nine diamonds denote M dwarfs with a FGK primary from Table~\ref{tab:fgk}, and the 25 pairs of circles denote M-M binaries from Table~\ref{tab:main}. For suspected unresolved binaries, the circle denotes the upper limit of the mass of the rotating component, with a dashed arrow indicating this point may be moved leftwards. When the masses of two components are known but it is unclear which is responsible for the observed rotation, the second option is denoted with an open circle, joined to the first with a dashed line. M-M pairs are colored based on their rotation rate: systems where both components are rapidly rotating ($P_{\rm rot} < 2$ days) are shown in shades of blue; systems where one component is rapidly rotating and the other has begun spindown are red; systems where both stars are slowly rotating are green. For the WD-M and FGK-M pairs, symbols are colored based on age: young ($< 1$ Gyr) systems are black, intermediate (1--3 Gyr) are gray, and those known to be old ($> 5$ Gyr) are white. Small, semi-transparent black crosses denote the single field M-dwarf sample from \citet{Newton2017}, with updated masses using \textit{Gaia} parallaxes. Rectangles show the binned Praesepe data from the right panel of Figure~\ref{fig:main}, with the dotted line showing the 90th percentile of Praesepe stars. The offset between the slowly-rotating sequence for early M dwarfs in Praesepe and the \citet{Newton2017} field sample is possibly surprising; if these stars stall at Praesepe-like periods for ages up to 2.7 Gyr \citep[][]{Curtis2020}, one might expect to observe a substantial number of field stars at these periods.}
    \label{fig:main2}
\end{figure*}

\needspace{6em}
\subsection{Discussion of FGK-M and WD-M pairs}
\label{sec:fgk}

While our sample of WD-M pairs is small, they showcase three potentially interesting regimes (Figure~\ref{fig:main2}). The two fully-convective M dwarfs that have spun down to rotation rates slower than 100 days are both likely older than 5 Gyr. The fully-convective M dwarf that is starting to spin down (with $P_{\rm rot} = 7.6$ days) is at least 800 Myr old, and likely in the range 1--3 Gyr. The rapidly-rotating, early M dwarf, G 68-34, is surprisingly old, with an age of at least 5.5 Gyr. \citet{Douglas2017} hypothesize that all the rapidly-rotating early M dwarfs in Praesepe and the Hyades may be unknown binaries; G 68-34 could fall into this category, with its rapid rotation persisting for gigayears due to interactions with a close, currently-unknown companion. This star should be investigated with radial velocity monitoring to establish whether it is a spectroscopic binary; if it is truly single, its existence would have important implications for our picture of M-dwarf spindown.

The ten FGK-M pairs were selected based on H\textalpha\ activity in the M-dwarf secondary and therefore represent systems in which the M dwarf has not yet spun down or where spindown is ongoing. Most intriguing of these pairs is HD 211472 \& G 232-62, where the 0.27M$_\odot$ M dwarf rotates with a period of 59 days but has an age similar to that of Praesepe. Such a long rotation period for a 600-Myr M dwarf is surprising; \citet{Douglas2019} did not identify any comparably-slow rotators in their K2 study of Praesepe. However, \citet{Reinhold2020} note that they typically measure rotation periods less than 44 days from K2 data, as longer periods would be longer than half the observation time span. It is therefore possible that a population of 0.3M$_\odot$ stars could be greatly spun down by Praesepe age, yet not contradict the observations of \citet{Douglas2019}. 

To investigate this hypothesis, we identify the single, \hbox{$<0.35$M$_\odot$} stars in Praesepe with H\textalpha\ measurements in \citet{Douglas2014}, selecting those with comparable or lesser H\textalpha\ emission to G 232-62 (EW $>$ -3\AA). We require that the stars are also identified as Praesepe members in \citet{Lodieu2019}, whose membership assessments benefit from recent \textit{Gaia} data. To be included in that catalog, the motion must be consistent with the cluster with a \hbox{p-value} above 0.99999; \citet{Lodieu2019} therefore consider all stars in their catalog to be bona fide Praesepe members. We then search for variability in the ZTF light curves. From this sample, we identify two Praesepe stars with clear, long rotation periods in ZTF: 2MASS J08394051+1918539, with a rotation period of 60 days and mass of 0.32M$_\odot$, and 2MASS J08380676+1934178, with a rotation period of 42 days and mass of 0.30M$_\odot$. Notably, \citet{Douglas2019} identified a 21-day rotation period for 2MASS J08380676+1934178 from the K2 data, confirming our suspicion that long rotation periods may be mistaken with their second harmonic given the limited observation baseline of K2. \citet{Rebull2017} identify a further two stars\footnote{2MASS J08372941+1841355 with mass 0.31M$_\odot$ \& \newline 2MASS J08420785+2211051 with mass 0.34M$_\odot$} in this sample as having `timescales' of 28 days based on K2 data, which they define as repeated patterns that change sufficiently over each period that they are unlikely to be true rotation periods. One explanation for such a timescale is that it is the second harmonic, which would yield rotation periods of 56 days. Periodogram peaks that are consistent with this hypothesis are present in the ZTF data of these two stars, but higher-precision photometry is needed to conclusively establish these periods.

In summary, there is indeed a population of 0.30--0.35M$_\odot$ stars rotating with periods of 40--60 days at the age of Praesepe. It may be that the lack of a well-defined slowly-rotating sequence at these masses in Figure~\ref{fig:main} is the result of the limited sensitivity of K2 to long rotation periods, and a slowly-rotating sequence is present at longer rotation periods. To further test this hypothesis, we recommend a systematic study of Praesepe M dwarfs with an instrument capable of a longer observing baseline than K2 and higher photometric precision than ZTF, such as the recently-commissioned Tierras Observatory \citep{GarciaMejia2020}.

Five other systems in the FGK-M sample have ages younger than 1 Gyr. These M dwarfs have diverse rotation periods: two with periods of a few days, two with periods near one day, and the lowest-mass star of the five with a period of 0.318 days. For the three stars with $P_{\rm rot} < 2$ days, our age estimates place them younger than Praesepe, and in each case, the observed rotation period is shorter than most Praesepe stars in the same mass bin (i.e., they fall below the Praesepe interquartile ranges in Figure~\ref{fig:main2}). This is consistent with our expectation that the rapidly-rotating stars in Praesepe have spun down slightly from their most rapidly-rotating state. We also consider the possibility that these stars should be adjusted to lower masses due to overluminosity at their younger ages, bringing them into closer alignment with the trend in Praesepe. The MIST evolutionary models \citep{Dotter2016, Paxton2011, Paxton2013, Paxton2015} indicate that the overluminosity of 0.2--0.3M$_\odot$ stars becomes negligible around ages of 200--300 Myr. \hbox{LP 99-392} and \hbox{LP 876-10} are therefore unlikely to be overluminous, meaning their masses are accurate. If \hbox{G 271-110} is indeed a member of the \textbeta\ Pic Moving Group, it is certainly overluminous; using MESA models, we find that an age of 24 Myr would reduce the mass estimate of G 271-110 by roughly a factor of 2.5.

The other two young M dwarfs possess masses near 0.3M$_\odot$ and rotation periods near \hbox{6 days}. LP 128-32 is unlikely to be affected by overluminosity, although the lithium-based age posterior for 2MASS J05363846+1117487 is consistent both with ages above and below 200 Myr. These stars are more slowly rotating than most comparable stars in Praesepe, although there are a small number of Praesepe stars with similar rotation periods and masses. These stars are also younger than Praesepe members, with age posteriors peaking at \hbox{370 Myr} for LP 128-32 and 210 Myr for 2MASS J05363846+1117487 based on the lithium abundance analysis of their primaries. This suggests that the rotation period dispersion observed in Praesepe for masses around 0.3M$_\odot$ may be carried forward from earlier ages. That is, these stars may never have spun up to rotation periods $P_{\rm rot} < 2$ days.

Four of the M dwarfs have ages $> 1$ Gyr (or three, neglecting GJ 166 C, which is a special case due to its white-dwarf companion). These stars are clustered in our mass-period diagram (Figure~\ref{fig:main2}), with masses around 0.2M$_\odot$ and rotation periods of a few days. While exact ages for stars in this category are difficult to establish---lithium abundance only provides a lower limit and the rotation periods of the primaries are longer than half of a TESS sector---our analysis suggests that each falls within the range of 1--3 Gyr. These points also appear near to \hbox{SCR J1107-3420 B}, the member of the WD-M sample with an age of 1--3 Gyr, depending on the initial-final mass relation for the white dwarf. This group of points is shifted to longer rotation periods than stars of similar mass in Praesepe, suggesting that the reservoir of rapid rotators has evolved from $P_{\rm rot} < 2$ days at 600 Myr to $2 < P_{\rm rot} < 10$ days at ages of 1--3 Gyr.

We note the caveat that 2MASS J22562702+7600101 is possibly an unresolved binary; while the 2.51-day modulation dominates the TESS light curve, we also see a small residual signal in the periodogram at 0.431 days. However, the star does not have an abnormal \textit{Gaia} RUWE, and we do not observe double lines in our $R=44000$ spectra or radial velocity variation in our three observations taken over a five-day interval (with RV uncertainties of roughly 50 m/s).

One might wonder whether the low-mass M dwarfs remaining in the rapidly-rotating reservoir at 1--3 Gyr are the majority, or if most stars with similar masses have already spun down to the slowly-rotating sequence by these ages. Neglecting GJ 166 C, our full sample of both inactive and active FGK-M pairs contains 12 M dwarfs with masses of 0.15--0.25M$_\odot$. Four of these stars are active: LP 876-10, with an age younger than Praesepe, and the three M dwarfs with ages of 1--3 Gyr. As discussed above, we do not think any of these stars are overluminous. The remaining eight stars are inactive. If one adopts the assumption from \citet{Medina2022} that star formation is uniform over the last 8 Gyr (which is motivated by the results of \citealt{Fantin2019}), we would obtain this ratio of inactive-to-active stars if 0.15--0.25M$_\odot$ M dwarfs typically become inactive around \hbox{1/3 $\times$ 8 Gyr = 2.7 Gyr}. Our sample is therefore consistent with the majority of 0.15--0.25M$_\odot$ M dwarfs remaining in the rapidly-rotating reservoir at 1--3 Gyr. Nevertheless, our sample size is fairly small.

\needspace{6em}
\subsection{Discussion of M-M pairs}
Our M-M pairs can be divided into three regimes, indicated by color in Figure~\ref{fig:main2}. Either both stars are rapidly rotating, both are slowly rotating, or one/both are intermediate rotators. While we do not observe a correlation between mass and rotation period for systems where both components are rapidly rotating, the transitioning systems (colored in red in Figure~\ref{fig:main2}) show positive slopes, indicating that the more massive component rotates more slowly in systems where spindown is ongoing. This provides validation of the mass-dependent picture of spindown that is inferred through observations of the M dwarfs in Praesepe; at a given age, more massive M dwarfs have spun down while less massive stars have not, and our M-M pairs represent two stars with different masses but the same age. A similar result was recently obtained by \citet{Medina2022} using galactic kinematics; these authors found that the average age of M dwarfs with $P_{\rm rot} < 10$ days is 2.3 Gyr for 0.1--0.2M$_\odot$ stars and 600 Myr for 0.2--0.3M$_\odot$ stars. 

However, we know that the time at which a fully-convective M dwarf transitions from the rapidly-rotating reservoir to the slowly-rotating sequence must be variable: stars such as 2MASS J02580617+2040016 (the inactive Hyad), \hbox{G 232-62} (the 59-day rotator in the FGK-M sample), and 2MASS J08394051+1918539 and 2MASS J08380676+1934178 (the slow rotators in Praesepe) appear to have spun down considerably by roughly 600 Myr. Other stars like LP 128-32 and 2MASS J05363846+1117487 in the FGK-M sample rotate at modest rotation periods of a few days at even younger ages, perhaps being the precursors to these 600 Myr-old stars. In light of this, the ordered behavior of the transitioning pairs may in fact be surprising: the variability in the epoch of spindown is unable to mask the strong correlation between mass and rotation rate.

Our M-M sample is influenced by some observational biases. To be included in this sample, we require a rotation period measurement for both components. It is easier to measure short rotation periods, as one observes more cycles over the same baseline. This effect becomes more pronounced for the TESS subsample, as long rotation periods cannot be identified at all. M dwarfs in the rapidly-rotating reservoir are therefore overrepresented in our sample. This would affect M dwarfs of all masses, and could explain why we observe a surprising number of early M dwarfs with $P_{\rm rot} < 2$ days, despite studies of clusters indicating that these early M dwarfs typically leave the rapidly-rotating reservoir at ages of a few 100 Myr (alternatively, these objects could be unresolved binaries, as mentioned in our discussion of G 68-34).

To probe the extent of this selection effect, we consider the FGK-M sample. Since we obtained rotation periods for all the M dwarfs that were H\textalpha-active in our larger sample of 31 FGK-M pairs (aside from the special case of GJ 166 C), these M dwarfs provide a relatively unbiased sample of the rotation periods of M dwarfs pre-spindown and during spindown (while young, overluminous stars may be overrepresented as this sample is brightness-limited, the majority of these stars are not overluminous; see discussion in Section~\ref{sec:fgk}). From this analysis, the overabundance of rapid rotators in the M-M sample is apparent: only a third of M dwarfs from the active FGK-M sample have $P_{\rm rot} < 2$ days, compared to roughly three quarters of the M-M sample over a similar mass range. The number of M-M pairs where the more massive component has a rotation period similar to stars in Praesepe should therefore not be interpreted as evidence for stalling in the more massive M dwarfs; it may be that many of these systems simply have ages comparable to Praesepe.

It is therefore more interesting to consider the subset of transitioning systems where the slower-rotating M dwarf rotates more slowly than most stars of similar mass in Praesepe; these systems are likely older than 600 Myr. Six of our M-M pairs fall into this category, decreasing to three if we neglect likely close binaries with uncertainty in the component masses. These three systems are \hbox{LHS 3808} \& \hbox{LHS 3809}, \hbox{LP 167-63} \& \hbox{LP 167-74}, and 2MASS J21005492-4131438 \& 2MASS J21010380-4114331.

LP 167-63 and LHS 3809 have masses of 0.14M$_\odot$ and therefore represent a lower-mass portion of the rapidly-rotating reservoir than probed by our other samples. These stars have rotation periods of 1.0 and 1.6 days, respectively, and their more massive components have spun down to rotation periods consistent with the inactive sequence in \citet{Newton2017}. All such low-mass M dwarfs observed in Praesepe are very rapidly rotating, with $P_{\rm rot} < 0.5$ days. The lowest-mass end of the reservoir therefore appears to also evolve to slower rotation periods with age, into the regime of 1--2 days. A sample of two stars is likely too small to probe the breadth of rotation periods in the reservoir at advanced ages; we note that other stars appear at these masses in the \citet{Newton2017} field sample with periods of up to 5 days (shown as crosses in our Figure~\ref{fig:main2}).

\subsection{Comparison with theoretical models}
\label{sec:garraffo}
\citet{Garraffo2018} posit a modified and physically-motivated version of MDM: stars still transition rapidly between fast-rotating and slow-rotating modes based on a change in their magnetic dynamo, but the epoch of this change is not stochastic. Rather, the evolution of a star's rotation is completely determined by two properties: the star's mass and its seed rotation rate. At a given mass, stars with more rapid initial rotation spend more time in the rapidly-rotating mode, while stars with slower seed rotation rates quickly shed their angular momentum.

The initial rotation rate is likely related to disk interactions and the epoch of disk dissipation; \citet{Rebull2018} note the general consensus that slowly-rotating stars in a young open cluster like the Pleiades are linked to long-lived primordial disks, while the rapidly-rotating population likely had their disks dissipate at younger ages. This distinction is the result of the disk providing a rotation lock that prevents the star from spinning up due to momentum conservation as it contracts. \citet{Roquette2021} note that the disk lifetime can be substantially influenced by the far-ultraviolet radiation from nearby massive stars, and therefore a star's initial rotation rate may be a proxy for its high-energy birth environment. Various works have found direct evidence supporting the disk-locking hypothesis using young ($<10$ Myr) star-forming regions (see Section 4.1 of \citealt{Rebull2018} for a review of this literature). In more recent work, \citet{Rebull2022} observe a pileup of disked mid-M dwarfs in 10--20 Myr clusters, sharply-peaked at rotation periods near 2 days; the location of this pileup is mass-dependent, with lower-mass M dwarfs rotating more rapidly.

\begin{figure}[t]
    \centering
    \includegraphics[width=0.9\columnwidth]{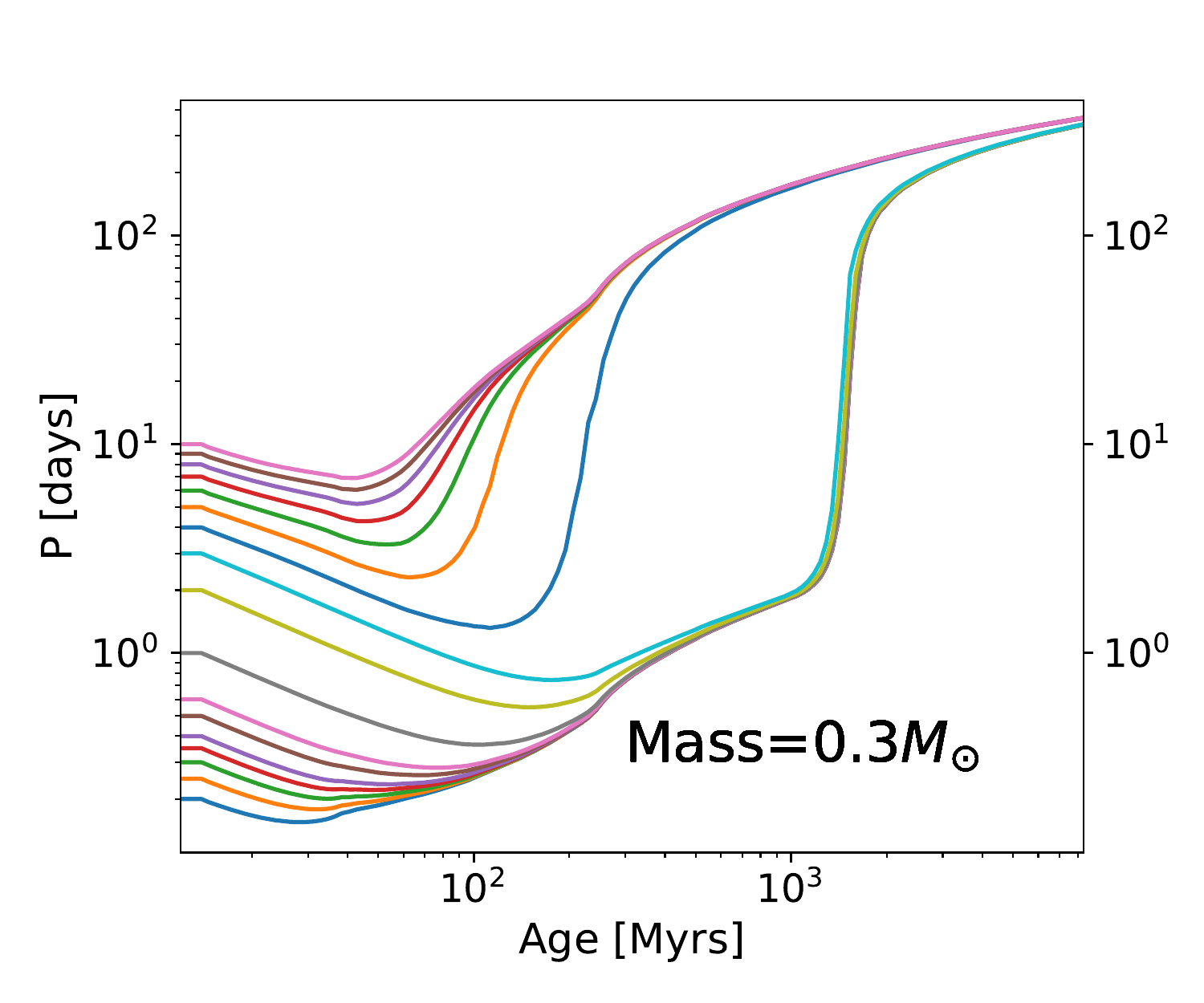}
    \vspace{-0.3cm}
    \caption{Rotation rate evolution of a 0.3M$_\odot$ main-sequence star with varying initial rotation rates (defined as the rotation in the immediate post-disk phase), calculated using the model of \citet{Garraffo2018} that considers changing magnetic complexity during spindown.}
    \label{fig:garraffo}
\end{figure}

The evolution predicted by the \citet{Garraffo2018} model for a 0.3M$_\odot$ star is shown in Figure~\ref{fig:garraffo}, calculated using their publicly-available evolution code.\footnote{\href{https://github.com/cgarraffo/Spin-down-model/}{https://github.com/cgarraffo/Spin-down-model/}} This model does not perfectly reproduce reality---for one, the final rotation periods are slower than observed in the field, and the code outputs non-physical results for models of lower-mass ($<$0.3M$_\odot$) M dwarfs---but nevertheless it produces some interesting predictions. In the context of this model, 0.3M$_\odot$ M dwarfs with an initial rotation rate less than a critical value of $P_{\rm rot} = 3$ days converge over a few 100 Myr onto the same rapidly-rotating rotation--age sequence, a process which could yield the rapidly-rotating reservoir in Praesepe. Stars with slower rotation rates in the immediate post-disk phase converge to a different, more slowly-rotating sequence by a few 100 Myr; this could explain spun-down young M dwarfs like 2MASS J02580617+2040016 and \hbox{G 232-62}. However, the prediction of two clean, converged sequences at such a young age does not match the Praesepe data, where a dispersion in rotation periods at a given mass is apparent. It therefore appears that the initial rotation cannot be the sole determinant of the spindown epoch for stars of a given mass in the fully-convective regime (or the convergence produced by the \citealt{Garraffo2018} model is premature). This may suggest some stochastic element, as posited by \citet{Brown2014}, or that additional parameters are necessary to establish a predictive model. Alternatively, see \citet{See2019}, who argue that this model is flawed and non-dipolar modes do not significantly contribute to stellar spindown.

Modified MDM is not the only model of stellar spindown, nor is it the only model to link the initial rotation rate to the epoch of spindown. For low-mass stars with radiative cores and convective envelopes, \citet{Denissenkov2010} argue that these components decouple and rotate differentially for stars with slower initial rotation, leading spindown to progress more rapidly; with more rapid initial rotation, the star rotates as a solid body and spindown is less efficient. However, this distinction cannot explain the fully-convective M dwarfs in this study that have spun down by 600-Myr ages. The \citet{Breimann2021} model (based on the torque law of \citealt{Matt2015}) is also highly-sensitive to initial rotation rates, with those authors finding a much better fit to Praesepe data when their model is initialized with the rotation rate distribution from Upper Sco instead of a uniform distribution. Their model does not require inefficient torques for rapid rotators, in contrast to the MDM models, although it is also unable to produce the observed slowly-rotating M dwarfs at Praesepe ages.

\section{Summary}
\label{sec:summary}
A model that explains M-dwarf spindown needs to be consistent with six key observables: i) the correlation between mass and rotation rate for the components of the M-M binary pairs; ii) the distribution of masses and rotation periods of M dwarfs in clusters like Praesepe, including the mass-dependent behavior of the rapidly-rotating reservoir; iii) the bimodality of M-dwarf rotation periods at field ages \citep[e.g.,][]{Newton2016, Newton2017, Newton2018}; iv) the fraction of fully-convective M dwarfs remaining in the rapidly-rotating mode at field ages (found to be 29\% in the volume-complete sample of \citealt{Medina2022}); v) the clustering at rotation periods of $2 < P_{\rm rot} < 10$ days for fully-convective M dwarfs with ages of 1--3 Gyr in our FGK-M and WD-M samples; and vi) the existence of fully-convective M dwarfs that have spun down by 600 Myr.

Spindown of these low-mass stars begins slowly, as evidenced by the evolution of the rotation periods of 0.2--0.3M$_\odot$ rapid rotators from 1 day at the age of Praesepe to a few days at 1--3 Gyr. Lower-mass stars experience a similar phenomenon but start with rotation periods around 0.2--0.5 days. At some point, the rate of change of the rotation rate increases dramatically (due to a change in magnetic dynamo, in the theories of \citealt{Brown2014} and \citealt{Garraffo2018}), over a timescale that must be short to create the bimodal appearance of the field sample. The time at which this transition occurs varies from star to star, and may be stochastic \citep[e.g.,][although this work does not posit a physical mechanism for the transition]{Brown2014} or linked to initial rotation rate \citep[e.g.,][]{Garraffo2018}. Based on the ages we find for H\textalpha-active, fully-convective M dwarfs in our FGK-M and WD-M samples, spindown likely occurs around 2--3 Gyr, although 2MASS J02580617+2040016 and \hbox{G 232-62} are examples of fully-convective M dwarfs that have spun down considerably by 600 Myr. These M dwarfs may be the descendants of stars like LP 128-32 and 2MASS J05363846+1117487, which are rotating in the $2 < P_{\rm rot} < 10$ regime by ages of a few 100 Myr. Such stars are predicted by the \citet{Garraffo2018} spindown model, where they represent systems whose disk dissipated later than average and hence experienced less spinup during contraction onto the main sequence.

Unresolved binaries complicate studies of spindown. Both H\textalpha\ emission and rapid rotation persist in old M dwarfs experiencing spin-orbit interactions, as we see in the GJ 1006 and GJ 1230 hierarchical triple systems. We find that G 68-34, a 0.46M$_\odot$ M dwarf, remains rapidly rotating and active at an age of $>5$ Gyr and posit an unresolved close binary companion as a possible explanation for this system. Such a companion would induce a radial velocity perturbation, and hence this hypothesis is testable with spectroscopic monitoring.

This work leveraged existing photometric time series to measure rotation periods of wide binary pairs. Future studies can improve upon our results by performing high-precision, long-baseline photometric monitoring of M dwarfs with the most power for discriminating between spindown models, such as M-M pairs where one component is H\textalpha-active and the other is not, or M-dwarf companions to early primaries known to be intermediate (1--3 Gyr) in age. Similar monitoring would be beneficial for Praesepe, establishing whether fully-convective, 600-Myr M dwarfs with rotation periods of 40--60 days are outliers, or if they form a slowly-rotating sequence.

\section*{Acknowledgements}

We thank Perry Berlind, Allyson Bieryla, Michael Calkins, Gilbert Esquerdo, Pascal Fortin, David Latham, Amber Medina, Jessica Mink, and Samuel Quinn for assistance in the collection and interpretation of the TRES spectra. This manuscript benefited from the comments of the anonymous referee.

E.P. is supported in part by an NSERC Postgraduate Scholarship. This work is made possible by a grant from the John Templeton Foundation. The opinions expressed in this publication are those of the authors and do not necessarily reflect the views of the John Templeton Foundation. This material is based upon work supported by the National Aeronautics and Space Administration under grants 80NSSC19K0635, 80NSSC19K1726, 80NSSC21K0367, 80NSSC22K0165, and 80NSSC22K0296 in support of the TESS Guest Investigator Program and grant 80NSSC18K0476 issued through the XRP program. This paper includes data collected by the TESS mission, which are publicly available from the Mikulski Archive for Space Telescopes (MAST). Funding for the TESS mission is provided by the NASA's Science Mission Directorate. The MEarth Project acknowledges funding from the David and Lucile Packard Fellowship for Science and Engineering, and the National Science Foundation under grants AST-0807690, AST-1109468, AST-1616624 and AST-1004488 (Alan T. Waterman Award).

This work has made use of data from the European Space Agency (ESA) mission
{\it Gaia} (\url{https://www.cosmos.esa.int/gaia}), processed by the {\it Gaia}
Data Processing and Analysis Consortium (DPAC,
\url{https://www.cosmos.esa.int/web/gaia/dpac/consortium}). Funding for the DPAC
has been provided by national institutions, in particular the institutions
participating in the {\it Gaia} Multilateral Agreement.

This work makes use of observations obtained with the Samuel Oschin 48-inch Telescope at the Palomar Observatory as part of the Zwicky
Transient Facility project. ZTF is supported by the National Science Foundation under Grant No. AST-1440341 and a
collaboration including Caltech, IPAC, the Weizmann Institute for Science, the Oskar Klein Center at Stockholm University, the
University of Maryland, the University of Washington, Deutsches Elektronen-Synchrotron and Humboldt University, Los Alamos
National Laboratories, the TANGO Consortium of Taiwan, the University of Wisconsin at Milwaukee, and Lawrence Berkeley
National Laboratories. Operations are conducted by COO, IPAC, and UW.

\facilities{CTIO:1.5m (CHIRON), FLWO:1.5m (TRES), Gaia, MEarth, PO:1.2m (ZTF), TESS}
\software{\texttt{Astropy} \citep{Astropy2013, Astropy2018}, \texttt{BAFFLES} \citep{StanfordMoore2020}, \texttt{BANYAN \textSigma} \citep{Gagne2018}, \texttt{eleanor} \citep{Feinstein2019}, \texttt{Lightkurve} \citep{lightkurve2018}, \texttt{Matplotlib} \citep{Hunter2007}, \texttt{NumPy} \citep{Harris2020}, \texttt{pandas} \citep{Reback2020}, \texttt{PyAstronomy} \citep{Czesla2019}, \texttt{PyMC3} \citep{Salvatier2016}, \texttt{SciPy} \citep{Scipy2020}, \texttt{stardate} \citep{Angus2019}}


\bibliography{mmbinary}{}
\bibliographystyle{aa_url}



\end{document}